\documentclass[amsmath,eqsecnum,preprint,pre,aps,showpacs]{revtex4}
\usepackage{amsmath}
\usepackage{graphicx}
\usepackage{epstopdf}
\usepackage{dcolumn}
\usepackage{bm}
\usepackage{float}

\begin{document}

\draft
\emph{}
\title{Equilibration  and aging of dense soft-sphere glass-forming liquids}
\author{Luis Enrique S\'anchez-D\'iaz,  Pedro Ram\'irez-Gonz\'alez, and
Magdaleno Medina-Noyola}

\address{Instituto de F\'{\i}sica {\sl ``Manuel Sandoval Vallarta"},
Universidad Aut\'{o}noma de San Luis Potos\'{\i}, \'{A}lvaro
Obreg\'{o}n 64, 78000 San Luis Potos\'{\i}, SLP, M\'{e}xico}

\date{\today}

\begin{abstract}
The recently-developed non-equilibrium extension of the
self-consistent generalized Langevin equation theory of
irreversible relaxation [Phys. Rev. E (2010) \textbf{82}, 061503;
ibid. 061504] is applied to the description of the irreversible
process of equilibration and aging of a glass-forming soft-sphere
liquid that follows a sudden temperature quench, within the
constraint that the local mean particle density remains uniform
and constant. For these particular conditions, this theory
describes the non-equilibrium evolution of the static structure
factor $S(k;t)$ and of the dynamic properties, such as the
self-intermediate scattering function $F_S(k,\tau;t)$, where
$\tau$ is the correlation \emph{delay} time and $t$ is the
\emph{evolution} or \emph{waiting} time after the quench. Specific
predictions are presented, for the deepest quench (to zero
temperature). The predicted evolution of the $\alpha$-relaxation
time $\tau_\alpha(t)$ as a function of $t$ allows us to define the
\emph{equilibration} time $t^{eq}(\phi)$, as the time after which
$\tau_\alpha(t)$ has attained its equilibrium value
$\tau_{\alpha}^{eq}(\phi)$.  It is predicted that both,
$t^{eq}(\phi)$ and $\tau_{\alpha}^{eq}(\phi)$, diverge as  $\phi
\to \phi^{(a)}$, where $\phi^{(a)}$ is  the hard-sphere
dynamic-arrest volume fraction $\phi^{(a)}\ (\approx0.582)$, thus
suggesting that the measurement of \emph{equilibrium} properties
at and above $\phi^{(a)}$ is experimentally impossible. The theory
also predicts that for fixed finite waiting times $t$, the plot of
$\tau_\alpha(t;\phi)$ as a function of $\phi$ exhibits two
regimes, corresponding to samples that have fully equilibrated
within this waiting time $(\phi\le \phi^{(c)}(t))$, and to samples
for which equilibration is not yet complete $(\phi\ge
\phi^{(c)}(t))$. The crossover volume fraction $\phi^{(c)}(t)$
increases with $t$ but saturates to the value $\phi^{(a)}$.

\end{abstract}


\maketitle

\section{Introduction.}\label{sectionI}

Classical and statistical thermodynamics deal with the equilibrium states of matter \cite{callen,mcquarrie}. Driving the system from one equilibrium state to another, however, involves the passage of the system through a sequence of instantaneous states that do not satisfy the conditions for thermodynamic equilibrium, and hence, constitute a non-equilibrium process \cite{degrootmazur,keizer}. The description of these processes fall outside the realm of classical and statistical thermodynamics, unless the sequence of non-equilibrium states do not depart appreciably from a sequence of equilibrium states. Such idealized process can be thought of as an infinite sequence of infinitesimally small changes in the driving control parameter, after each of which the system is given sufficient time to equilibrate. This so-called quasistatic process is an excellent representation of real process when the equilibration times of the system are sufficiently short. However, when the equilibration kinetics is very slow, virtually any change will involve intrinsically non-equilibrium states whose fundamental understanding must unavoidably be done from the perspective of a non-equilibrium theory \cite{casasvazquez0}.

These concepts become particularly relevant for the description of the slow dynamics of metastable glass-forming liquids in the vicinity of the glass transition \cite{angell,debenedetti}. It is well known that the decay time of the slowest relaxation processes (the so-called $\alpha$-relaxation time $\tau_\alpha$) increases without bound as the temperature $T$ is lowered below the glass transition temperature $T^{(g)}$. It is then natural to think that the \emph{equilibration} time of the system must also increase accordingly. To be more precise, let us imagine that a glass-forming liquid, initially at an arbitrary temperature $T^{(i)}$, is suddenly cooled at time $t=0$ to a final temperature $T$, after which it is allowed to evolve spontaneously toward its thermodynamic equilibrium state. Imagine that we then monitor its $\alpha$-relaxation time $\tau_\alpha (t)$ as a function of the evolution or ``waiting'' time  $t$ elapsed after the quench. We say that the system has equilibrated when  $\tau_\alpha (t)$ reaches the plateau that defines its final equilibrium value  $\tau_\alpha^{eq}(T)$, which must only depend on the final temperature $T$. The beginning of this plateau occurs at a certain value of the waiting time $t$, that we refer to as the \emph{equilibration} time  $t^{eq}(T)$; this equilibration time must also depend on the final temperature $T$.

There are strong indications, from recent computer simulation experiments \cite{gabriel,kimsaito}, that in the metastable regime these two characteristic times, $\tau_\alpha^{eq}(T)$ and  $t^{eq}(T)$, are related to each other as  $t^{eq}(T)\propto [\tau_\alpha^{eq}(T)]^\eta$, with an exponent  $\eta > 1$ (more specifically, $\eta\approx 1.5$ \cite{gabriel,kimsaito}). This implies that in order to measure the actual equilibrium value $\tau_\alpha^{eq}(T)$ we have to wait, before starting the measurement of  $\tau_\alpha^{eq}(T)$, for an  equilibration time $t^{eq}(T)$ that will increase faster than $\tau_\alpha^{eq}(T)$ itself. This poses an obvious practical problem for the measurement of  $\tau_\alpha^{eq}(T)$ when the temperature $T$ approaches the glass transition temperature $T^{(g)}$, since sooner or later we shall be unable to wait this required equilibration time. This situation then implies that it is  impossible to discard a scenario in which the \emph{equilibrium} $\alpha$-relaxation time $\tau_\alpha^{eq}(T)$ diverges at a singular temperature $T^{(a)}$, since the equilibration time $t^{eq}(T)$ needed to observe this divergence will also diverge at that temperature, i.e., it will be impossible to equilibrate the system at a final temperature near or below $T^{(a)}$ within experimental waiting times. Of course, a measurement carried out at a finite $t$, will always report a result for $\tau_\alpha$, but this result will correspond to $\tau_\alpha (t)$, the non-equilibrium value of the $\alpha$-relaxation time registered at that waiting time $t$. Thus, the analysis of these experimental measurements cannot be based on the postulate that the system has reached equilibrium; instead, one needs to interpret these experiments in the framework of a quantitative theory of slowly-relaxing non-equilibrium processes.

Until recently, however, no \emph{quantitative, first-principles} theory had been
developed and applied to describe the slow \emph{non-equilibrium} relaxation of \emph{structural}
glass-forming atomic or colloidal liquids. About a decade ago Latz
\cite{latz} attempted to extend the conventional mode coupling
theory (MCT) of the ideal glass transition
\cite{goetze1,goetze2,goetze3,goetze4}, to describe the aging of
suddenly quenched glass forming liquids. A major aspect of his work
involved the generalization to non-equilibrium conditions of the
conventional equilibrium projection operator approach \cite{berne}
to derive the corresponding memory function equations in which the
mode coupling approximations could be introduced. Similarly, De
Gregorio et al. \cite{degregorio} discussed time-translational
invariance and the fluctuation-dissipation theorem in the context of
the description of slow dynamics in system out of equilibrium but
close to dynamical arrest. They also proposed extensions of
approximations long known within MCT. Unfortunately, in neither of
these theoretical efforts, quantitative predictions were
presented that could be contrasted with experimental or simulated
results in specific model systems of \emph{structural} glass-formers.

In an independent but similarly-aimed effort, on the other hand, the self-consistent generalized Langevin equation
(SCGLE) theory of colloid dynamics \cite{scgle1,scgle2,marco1,marco2} and of dynamic arrest \cite{rmf,todos1,todos2,rigo2,luis1} has recently been extended to describe the (non-equilibrium) spatially non-uniform and temporally non-stationary  evolution of
glass-forming colloidal liquids. Such an extension was introduced and described in
detail in Ref. \cite{nescgle1}, and will be referred to as the \emph{non-equilibrium} self-consistent
generalized Langevin equation (NE-SCGLE) theory. As one can imagine, the number and variety of the phenomena that
could be studied with this new theory may be enormous, and to start its systematic application we must focus on simple classes of physically relevant conditions. Thus, as a first simple illustrative application, this theory was applied in Ref. \cite{nescgle2} to a model colloidal liquid with hard-sphere plus short-ranged attractive interactions, suddenly quenched to an attractive glass state.

The aim of the present work is to start a systematic exploration of the scenario predicted by this theory when  applied to the simplest irreversible processes, in the simplest and best-defined model system. In the present case we refer to the irreversible isochoric evolution of a glass-forming liquid of particles interacting through purely repulsive soft-sphere interactions, initially at a fluid-like state, whose temperature is suddenly quenched to a final value $T^{(f)}=0$, at which the expected equilibrium state is that of a hard-sphere liquid at volume fraction  $\phi$. Such process mimics the spontaneous search for the equilibrium state of this hard-sphere liquid, driven to non-equilibrium conditions by some perturbation (shear, for example \cite{brambilla,elmasri}) which ceases at a time $t=0$. One possibility is that the system will recover its equilibrium state within an equilibration time $t^{eq}(\phi)$ that depends on the fixed volume fraction $\phi$. The other possibility is that the system ages forever in the process of becoming a glass. The application of the NE-SCGLE theory to these irreversible processes results in a well-defined scenario of the spontaneous non-equilibrium response of the system, whose main features are explained and illustrated in this paper.

In the following section we provide a brief summary of the non-equilibrium self-consistent
generalized Langevin equation theory, appropriately written to describe the equilibration of a monocomponent glass-forming liquid constrained to remain spatially uniform. Section \ref{sectionIII} defines the specific model to which this theory will be applied, discusses the strategy of solution of the resulting equations, and illustrates the main features of the results.  Section \ref{sectionIV} presents the scenario predicted by the NE-SCGLE theory for the first possibility mentioned above, namely, that the system is able to reach its thermodynamic equilibrium state. In this case we find  that the equilibrium $\alpha$-relaxation time $\tau_\alpha^{eq}(\phi)$, and the equilibration time $t^{eq}(\phi)$ needed to reach it, will remain finite for volume fractions smaller than a critical value $\phi^{(a)}$, but that both characteristic times will diverge as  $\phi$ approaches this dynamic-arrest volume fraction $\phi^{(a)}\approx 0.582$, and will remain infinite for $\phi\ge \phi^{(a)}$. Although it is intrinsically impossible to witness the actual predicted divergence, the theory makes distinct predictions regarding the transient non-equilibrium evolution occurring within experimentally-reasonable waiting times $t$.

In Sect. \ref{sectionV} we analyze the complementary regime, $\phi \ge \phi^{(a)}$, in which the system, rather than reaching equilibrium within finite waiting times, is predicted to age forever. In this regime we find that
the long-time asymptotic limit of  $S(k;t)$ will no longer be the expected equilibrium static structure factor $S^{(eq)}(k)$, but another, non-equilibrium but well-defined, static structure factor, that we denote as $S^{(a)}(k)$, and which depends on the protocol of the quench. Furthermore, contrary to the kinetics of the equilibration process, in which  $S(k;t)$ approaches $S^{(eq)}(k)$ in an exponential-like fashion, this time the decay of $S(k;t)$ to its asymptotic value $S^{(a)}(k)$ follows a much slower power law.

In section \ref{sectionVI} we put together the two regimes just described, in an integrated picture, which outlines the predicted scenario for the crossover from equilibration to aging. There we find that the discontinuous and singular behavior underlying the previous scenario is intrinsically unobservable, due to the finiteness of the experimental measurements, which constraints the observations to finite time windows. This practical but fundamental limitation converts the discontinuous dynamic arrest transition into a blurred crossover, strongly dependent on the protocol of the experiment and of the measurements.

The main purpose of the present paper is to explain in sufficient detail the methodological aspects of the application of the theory, so as to serve as a reliable reference for the eventual application of this non-equilibrium theory to the same system but with different non-equilibrium processes (e.g., different quench protocols), or in general to different systems and processes. Thus, we shall not report here the results of the systematic quantitative comparison of the scenario explained here with available specific simulations or experiments, which are being reported separately. Thus, the final section of the paper briefly refers to the main features of those comparisons, and discusses possible directions for further work.

\section{Review of the NE-SCGLE theory.}\label{sectionII}

Let us mention that the referred non-equilibrium self-consistent
generalized Langevin equation (NE-SCGLE) theory derives from a
non-equilibrium extension of Onsager's theory of thermal
fluctuations \cite{nescgle1},  and it consists of the time evolution
equations for the mean value $\overline{n}(\textbf{r},t)$ and for
the covariance $\sigma(\textbf{r},\textbf{r}';t)\equiv
\overline{\delta n (\textbf{r},t)\delta n (\textbf{r}',t)}$ of the
fluctuations $\delta n(\textbf{r},t) = n(\textbf{r},t)-
\overline{n}(\textbf{r},t)$ of the local concentration profile
$n(\textbf{r},t)$ of a colloidal liquid. These two equations are
coupled, through a local mobility function $b(\textbf{r},t)$, with
the two-time correlation function
$C(\textbf{r},\textbf{r}';t,t')\equiv \overline{\delta n
(\textbf{r},t)\delta n (\textbf{r}',t')}$. A set of well-defined
approximations on the memory function of
$C(\textbf{r},\textbf{r}';t,t')$, detailed in Ref. \cite{nescgle1},
results in the referred NE-SCGLE theory.

As discussed in Ref. \cite{nescgle1}, for given interparticle
interactions and applied external fields, the NE-SCGLE
self-consistent theory is in principle able to describe the
evolution of a strongly correlated liquid from an initial state with
arbitrary mean and covariance $\overline{n}^0(\textbf{r})$ and
$\sigma^0(\textbf{r},\textbf{r}')$, towards its equilibrium state
characterized by the equilibrium local concentration profile
$\overline{n}^{eq}(\textbf{r})$ and equilibrium covariance
$\sigma^{eq}(\textbf{r},\textbf{r}')$. These equations are in
principle quite general, and contain well known theories as
particular limits. For example, ignoring certain memory function
effects, the evolution equation for the mean profile
$\overline{n}(\textbf{r},t)$ becomes the fundamental equation of
dynamic density functional theory \cite{tarazona1}, whereas the
``conventional" equilibrium SCGLE theory \cite{todos2} (analogous in
most senses to MCT \cite{goetze1}) is recovered when full
equilibration is assumed and spatial heterogeneities are suppressed.
The NE-SCGLE theory, however, provides a much more general
theoretical framework, which in principle describes the spatially
heterogeneous and temporally non-stationary evolution of a liquid
toward its ordinary stable thermodynamic equilibrium state. This
state, however, will become unreachable if well-defined dynamic
arrest conditions arise along the equilibration pathway, in which
case the system evolves towards a distinct and  predictable
dynamically arrested state through an evolution process that
involves aging as an essential feature.

To start the systematic application of this general theory to more specific phenomena we must focus on a simple class of physical conditions. Thus, let us consider the irreversible evolution of the
structure and dynamics of a system \emph{constrained} to suffer a
programmed process of spatially \emph{homogeneous} compression or expansion
(and/or of cooling or heating). Under these conditions, rather than
solving the time-evolution equation for $\overline{n}({\bf r};t)$, we assume
that the system is constrained to remain \emph{spatially uniform},
$\overline{n}({\bf r};t)=\overline{n}(t)$, according to a
\emph{prescribed} time-dependence $\overline{n}(t)$ of the uniform
bulk concentration and/or to a prescribed uniform time-dependent
temperature $T(t)$. Among the many possible programmed protocols ($\overline{n}(t)$,
$T(t)$) that one could devise to drive or to prepare the system, in this paper we restrict ourselves to one of the simplest and most fundamental
protocols, which corresponds to the limit in which the system,
initially at an equilibrium state determined by initial values of
the control parameters, $(\overline{n}^{(i)},T^{(i)})$, must adjust
itself in response to a sudden and instantaneous change of these
control parameters to new values $(\overline{n}^{(f)},T^{(f)})$,
according to the ``program" $\overline{n}(t) =
\overline{n}^{(i)}\theta (-t)+\overline{n}^{(f)}\theta (t)$ and
$T(t) = T^{(i)}\theta (-t)+T^{(f)}\theta (t)$, with $\theta (t)$
being Heavyside's step function. Furthermore, just like in the first illustrative example described in Ref.
\cite{nescgle2}, here we shall also restrict ourselves to  the description of an even simpler subclass
of irreversible processes, namely, the isochoric cooling or heating of the system, in which its number density is constrained to remain  constant, i.e.,  $\overline{n}(t)=\overline{n}^{(i)}=\overline{n}^{(f)}=\overline{n}$, while the temperature $T(t)$ changes abruptly from its initial constant value $T^{(i)}$ to a final constant value $T^{(f)}$ at $t=0$.

Under conditions of spatial uniformity,
$C(\textbf{r},\textbf{r}';t,t')$ can be written as
\begin{equation}
C(\mid\textbf{r}-\textbf{r}'\mid,t'-t;\ t)= \frac{\overline{n}}{(2\pi)^3}\int d
\textbf{k}  \exp [-i\textbf{k}\cdot (\textbf{r}-\textbf{r}')]
F(k,\tau;t),
\end{equation}
with $\tau \equiv (t'-t) \ge 0$, and where $F(k,\tau;t)$ is the \emph{t}-evolving
non-equilibrium intermediate scattering function (NE-ISF). Similarly, the
covariance  $\sigma(\textbf{r},\textbf{r}';t)$ can be written as
\begin{equation}
\sigma(\mid\textbf{r}-\textbf{r}'\mid;t)= \frac{\overline{n}}{(2\pi)^3}\int d
\textbf{k}  \exp [-i\textbf{k}\cdot (\textbf{r}-\textbf{r}')]
S(k;t)
\label{ftsigma}
\end{equation}
with $S(k;t)\equiv F(k,\tau=0;t)$ being the time-evolving static structure factor. Under these conditions, the NE-SCGLE theory determines that the time-evolution
equation for the covariance (Eq. (2.11) of Ref. \cite{nescgle2}) may be written as an equation for $S(k;t)$ which, for $t>0$, reads
\begin{equation}
\frac{\partial S(k;t)}{\partial t} = -2k^2 D^0
b(t)\overline{n}^{(f)}\mathcal{E}^{(f)}(k) \left[S(k;t)
-1/\overline{n}\mathcal{E}^{(f)}(k)\right]. \label{relsigmadif2pp}
\end{equation}
In this equation the function
$\mathcal{E}^{(f)}(k)=\mathcal{E}(k;\overline{n},T^{(f)})$ is the
Fourier transform (FT) of the functional derivative
$\mathcal{E}[\mid\textbf{r}-\textbf{r}'\mid;n,T] \equiv \left[
{\delta \beta\mu [{\bf r};n]}/{\delta n({\bf r}')}\right]$,
evaluated at $n({\bf r})=\overline{n}$ and $T=T^{(f)}$.  As
discussed in Refs. \cite{nescgle1,nescgle2}, this thermodynamic
object embodies the information, assumed known, of the chemical
equation of state, i.e., of the functional dependence of the
electrochemical potential $\mu [{\bf r};n]$ on the number density
profile $n({\bf r})$.

The solution of this equation, for arbitrary initial condition $S(k;t=0)=S^{(i)}(k)$, can be written as
\begin{equation}
S(k;t)=S^{(i)}(k)e^{-\alpha (k)u(t)}+
[\overline{n}\mathcal{E}^{(f)}(k)]^{-1}\left(1-e^{-\alpha (k)
u(t)}\right), \label{solsigmadkt}
\end{equation}
with
\begin{equation}
\alpha (k) \equiv 2k^2D^0\overline{n}\mathcal{E}^{(f)}(k),
\label{alphadk}
\end{equation}
and with
\begin{equation}
u(t) \equiv \int_0^t b(t')dt'.
\label{udt}
\end{equation}

In the equations above, the time-evolving mobility $b(t)$ is defined as $b(t)\equiv
D_L(t)/D^0$, with $D^0$ being the short-time self-diffusion
coefficient  and $D_L(t)$ the long-time self-diffusion coefficient
at evolution time $t$. As
explained in Refs. \cite{nescgle1} and \cite{nescgle2}, the equation
\begin{equation}
b(t)= [1+\int_0^{\infty}
d\tau\Delta{\zeta}^*(\tau; t)]^{-1}
\label{bdt}
\end{equation}
relates $b(t)$ with the $t$-evolving,
$\tau$-dependent friction coefficient $\Delta{\zeta}^*(\tau; t)$
given approximately by
\begin{equation}
\begin{split}
  \Delta \zeta^* (\tau; t)= \frac{D_0}{24 \pi
^{3}\overline{n}}
 \int d {\bf k}\ k^2 \left[\frac{ S(k;
t)-1}{S(k; t)}\right]^2  \\ \times F(k,\tau; t)F_S(k,\tau; t).
\end{split}
\label{dzdtquench}
\end{equation}
Thus, the presence of $b(t)$ in Eq. (\ref{udt}) couples the formal solution for
$S(k;t)$ in Eq. (\ref{solsigmadkt}) with the solution of the
non-equilibrium version of the SCGLE equations for the collective
and self NE-ISFs $F(k,\tau; t)$ and $F_S(k,z; t)$. These equations
are written, in terms of the Laplace transforms (LT) $F(k,z; t)$ and
$F_S(k,\tau; t)$, as
\begin{gather}\label{fluctquench}
 F(k,z; t) = \frac{S(k; t)}{z+\frac{k^2D^0 S^{-1}(k;
t)}{1+\lambda (k)\ \Delta \zeta^*(z; t)}},
\end{gather}
and
\begin{gather}\label{fluctsquench}
 F_S(k,z; t) = \frac{1}{z+\frac{k^2D^0 }{1+\lambda (k)\ \Delta
\zeta^*(z; t)}},
\end{gather}
with $\lambda (k)$ being a phenomenological ``interpolating
function" \cite{todos2}, given by
\begin{equation}
\lambda (k)=1/[1+( k/k_{c})
^{2}],
\label{lambdadk}
\end{equation}
with $k_c=1.305\times k_{max}(t)$, where  $k_{max}(t)$ is the position of the main peak of $S(k; t)$ (in practice, however, $k_c\approx 1.305(2\pi/\sigma$) \cite{gabriel}). The simultaneous solution
of Ecs. (\ref{relsigmadif2pp})-(\ref{fluctsquench}) above, constitute
the NE-SCGLE description of the spontaneous evolution of the
structure and dynamics of an \emph{instantaneously} and
\emph{homogeneously} quenched liquid.

Of course, one important aspect of this analysis refers to the
possibility that along the process the system happens to reach the condition of dynamic arrest. For the discussion of this important aspect it is useful to consider the long-$\tau$ (or small $z$) asymptotic stationary solutions of Eqs. (\ref{fluctquench})-(\ref{dzdtquench}), the so-called non-ergodicity parameters, which are given by \cite{nescgle1}
\begin{equation}
f(k;t)\equiv \lim_{\tau\to\infty} \frac{F(k,\tau;t)}{S(k)} = \frac
{\lambda(k;t)S(k;t)}{\lambda(k;t)S(k;t)+k^2\gamma(t)} \label{fdkinf}
\end{equation}
and
\begin{equation}
f_S(k;t)\equiv \lim_{\tau\to\infty} F_S(k,\tau;t) = \frac
{\lambda(k;t)}{\lambda(k;t)+k^2\gamma(t)}, \label{fdksinf}
\end{equation}
where the $t$-dependent squared localization length $\gamma (t)$ is the solution
of
\begin{equation}
\frac{1}{\gamma(t)} =
\frac{1}{6\pi^{2}\overline{n}^{(f)}}\int_{0}^{\infty }
dkk^4\frac{\left[S(k;t)-1\right] ^{2}\lambda^2 (k;t)}{\left[\lambda
(k;t)S(k;t) + k^2\gamma(t)\right]\left[\lambda (k;t) +
k^2\gamma(t)\right]}. \label{nep5pp}
\end{equation}
Notice also that these equations are the
non-equilibrium extension of the corresponding results of the
equilibrium SCGLE theory (referred to as the ``bifurcation
equations" in the context of MCT \cite{goetze1}), and their derivation from Eqs.
(\ref{dzdtquench})- (\ref{fluctsquench}) follows the same arguments
as in the equilibrium case \cite{scgle2}. The solution ${\gamma(t)}$ of  Eq. (\ref{nep5pp}) and the mobility
$b(t)$ constitute two complementary dynamic order parameters, in the
sense that if ${\gamma(t)}$ is finite (or $b(t)=0$), then the system
must be considered dynamically arrested at that waiting time $t$,
whereas if $\gamma(t)$ is infinite, then the particles retain a
finite mobility, $b(t)>0$, and the instantaneous state of the system
is ergodic or fluid-like.

We recall that the first
relevant application of Eq. (\ref{nep5pp}) is the determination of
the \emph{equilibrium} dynamic arrest diagram in control-parameter
space (which, in the present case, is the density-temperature plane
$(n,T)$). This diagram determines the region of fluid-like states,
for which the solution $\gamma^{eq}(n,T)$ (of Eq. (\ref{nep5pp}),
with $S(k;t)=S^{eq}(k;n,T)$) is infinite. The complementary region
contains the dynamically-arrested states, for which
$\gamma^{eq}(n,T)$ is finite. The borderline between these two
regions is the dynamic arrest transition line. Due to the complementarity of the dynamic order parameters $\gamma(t)$ and $b(t)$, this curve is also the borderline between the region where the mobility $b(t)$ will reach its equilibrium value, $\lim_{t\to\infty}b(t)=b^{eq}(n,T)\ge0$, and the region of arrested states, where $\lim_{t\to\infty}b(t)=0$. Thus, since $b^{eq}(n,T)=D^*(n,T)\equiv D_L(n,T)/D^0$, where $D_L(n,T)$ is the equilibrium long-time self-diffusion coefficient at the point $(n,T)$, this line is also the iso-diffusivity curve corresponding to $D^*=0$.

\section{General features of the solution and a specific illustration.}\label{sectionIII}

Let us now discuss some general features of the solution of the
NE-SCGLE equations just presented. This discussion  has a
general character, but for the sake of clarity we shall illustrate the main concepts in the context of one specific application. Thus, consider a mono-component fluid of soft spheres
of diameter $\sigma$, whose particles interact
through the truncated Lennard-Jones (TLJ) pair potential that vanishes for $r\ge \sigma$, but which for $r\le \sigma$ is given, in units of the thermal energy
$k_BT=\beta^{-1}$, by
\begin{equation}
\beta u(r)= \epsilon\left[
\left( \frac{\sigma}{r}\right)^{2\nu} -2\left(
\frac{\sigma}{r}\right)^{\nu}+1 \right]. \label{truncatedlj}
\end{equation}
The state space of
this system is spanned by the volume fraction $\phi = \pi
\overline{n} \sigma^3/6$ and the reduced temperature $T^*\equiv k_BT
/\epsilon$.

\subsection{Thermodynamic framework: local curvature of the free energy surface.}\label{subsectionIII.1}

In order to apply Eqs. (\ref{relsigmadif2pp})-(\ref{fluctsquench}) to this model system, we first need to determine its thermodynamic property
$\mathcal{E}^{(f)}(k)$. As indicated above, this is  the
Fourier transform  of the functional derivative
$\mathcal{E}[\mid\textbf{r}-\textbf{r}'\mid;n,T] \equiv \left[
{\delta \beta\mu [{\bf r};n]}/{\delta n({\bf r}')}\right]$, which can also be written as
$\mathcal{E}[\mid\textbf{r}-\textbf{r}'\mid;n,T] = \delta (\textbf{r}-\textbf{r}')/\overline{n}- c(\mid\textbf{r}-\textbf{r}'\mid;n,T)$, with $c(r;n,T)$ being the ordinary direct correlation function \cite{mcquarrie}. This is an intrinsically thermodynamic property, related with the  \emph{equilibrium} static structure factor $S^{(eq)}(k;\overline{n},T)$ by the Ornstein-Zernike (OZ) equation, which in Fourier space reads $\overline{n}\mathcal{E}(k;\overline{n},T)S^{(eq)}(k;\overline{n},T)=1$. The OZ equation is the basis for the construction of the approximate integral equations of the \emph{equilibrium} statistical thermodynamics of liquids \cite{mcquarrie}. In fact, we shall employ one such approximation to determine $\overline{n}\mathcal{E}(k;\overline{n},T)$ for our soft-sphere system. This approximation, explained in detail in the appendix of Ref. \cite{soft1} and denoted as PY/VW, is based on the Percus-Yevick approximation \cite{percusyevick} within the Verlet-Weis correction \cite{verletweiss} for the hard sphere system, complemented by the treatment of soft-core potentials introduced by Verlet and Weis themselves \cite{verletweiss}.

Let us emphasize that for the present purpose, approximations such as these must be regarded solely as a practical and approximate mean to determine the thermodynamic property $\overline{n}\mathcal{E}(k;\overline{n},T)$, which is essentially the local curvature of the free energy surface at the state point $(\overline{n},T)$ \cite{nescgle1,evans}. This property directly determines the \emph{equilibrium} structure factor $S^{(eq)}(k;\overline{n},T)$  through the equilibrium relationship $\overline{n}\mathcal{E}(k;\overline{n},T)S^{(eq)}(k;\overline{n},T)=1$, and in practice we actually use this relationship to determine $\overline{n}\mathcal{E}(k;\overline{n},T)$. The main message of Eq. (\ref{relsigmadif2pp}), however, is that the experimentally observable, non-equilibrium, static structure factor $S(k;t)$ is not determined by any Ornstein-Zernike equilibrium condition, but by Eq. (\ref{relsigmadif2pp}) itself, with the thermodynamic property $\overline{n}\mathcal{E}(k;\overline{n},T)$  driving the non-equilibrium evolution in the manner indicated by its explicit appearance in this equation.

\subsection{Thermodynamic equilibrium vs. dynamically arrested states.}\label{subsectionIII.2}

In what follows, we are interested in studying the scenario revealed by the solution $S(k;t)$ of  Eq. (\ref{relsigmadif2pp}), for the process of isochoric equilibration (or lack of equilibration) of the static structure of a system subjected to a temperature control protocol $T(t) = T^{(i)}\theta (-t)+T^{(f)}\theta (t)$, corresponding to a an instantaneous temperature quench to a final temperature $T^{(f)}$ denoted simply as $T$. Thus, the system is assumed to be prepared at an initial equilibrium homogeneous state characterized by a bulk particle number density $\overline{n}$ and temperature $T^{(i)}$, at which its initial static structure factor is $S(k;t=0)=S^{(i)}(k)$. Upon suddenly changing the temperature of this system to the new value $T$, one normally expects that the system will reach full thermodynamic equilibrium, i.e., that the long-time asymptotic limit of $S(k;t)$ will be  the equilibrium static structure factor $S^{(eq)}(k;\overline{n},T)=1/\overline{n}\mathcal{E}(k;\overline{n},T)$. According to Eq. (\ref{relsigmadif2pp}), reaching this value is also a sufficient condition for $S(k;t)$ to reach a \emph{stationary} state.

According to the same equation, however, this is \emph{not a necessary} condition for the stationarity of $S(k;t)$, which could also be attained  if $\lim_{t \to \infty} b(t)  =0$, even in the absence of thermodynamic equilibrium (i.e., even if $\lim_{t \to \infty} S(k;t)\ne 1/\overline{n}\mathcal{E}(k;\overline{n},T)$). If the long-time stationary state attained is the thermodynamic equilibrium state, we say that the system is ergodic at the point $(\overline{n},T)$. The second condition, in contrast, corresponds to dynamically arrested states, in which the long-time asymptotic limit of $S(k;t)$ might differ from the expected thermodynamic equilibrium value $S^{(eq)}(k;\overline{n},T)=1/\overline{n}\mathcal{E}(k;\overline{n},T)$. Clearly, these are two mutually exclusive and fundamentally different classes of possible stationary states which can only be distinguished if we know the long-time limit of $b(t)$. This is, however, not a thermodynamic property, and hence, the discrimination of the ergodic or non-ergodic nature of the state point $(\overline{n},T)$ must be based on a dynamic or transport theory that allows the determination of $b(t)$.

One such theory is precisely the SCGLE theory: to decide if the long-time stationary state corresponding to the point $(\overline{n},T)$ will be an ergodic or an arrested state one can use the equilibrium static structure factor  $S^{(eq)}(k;\overline{n},T)$ in Eq. (\ref{nep5pp}) to calculate
$\gamma^{(eq)}(\overline{n},T)$. If the solution is infinite, we say that the asymptotic stationary state is ergodic, and hence, that at the point $(\overline{n},T)$ the system will be able to reach its thermodynamic equilibrium state without impediment, so  that $\lim_{t \to \infty} S(k;t) = 1/\overline{n}\mathcal{E}(k;\overline{n},T)$. On the other hand, if the solution for $\gamma^{(eq)}(\overline{n},T)$ turns out to be finite, this means that the system will become dynamically arrested, and that the long-time limit of $S(k;t)$ at the point $(\overline{n},T)$ will not necessarily be its thermodynamic equilibrium value $S^{(eq)}(k;\overline{n},T)=1/\overline{n}\mathcal{E}(k;\overline{n},T)$. Instead,  we shall have that $\lim_{t \to \infty} S(k;t) =  S^{(a)}(k)$, with a truly non-equilibrium structure factor $S^{(a)}(k)$, different from $S^{(eq)}(k;\overline{n},T)$, and obtained as an alternative stationary solution of  Eq. (\ref{relsigmadif2pp}). In this manner, by calculating $\gamma^{(eq)}(\overline{n},T)$ at all state points $(\overline{n},T)$ one can scan the state space to determine the region of dynamically arrested states of the system.

\begin{figure}
\includegraphics[scale=.27]{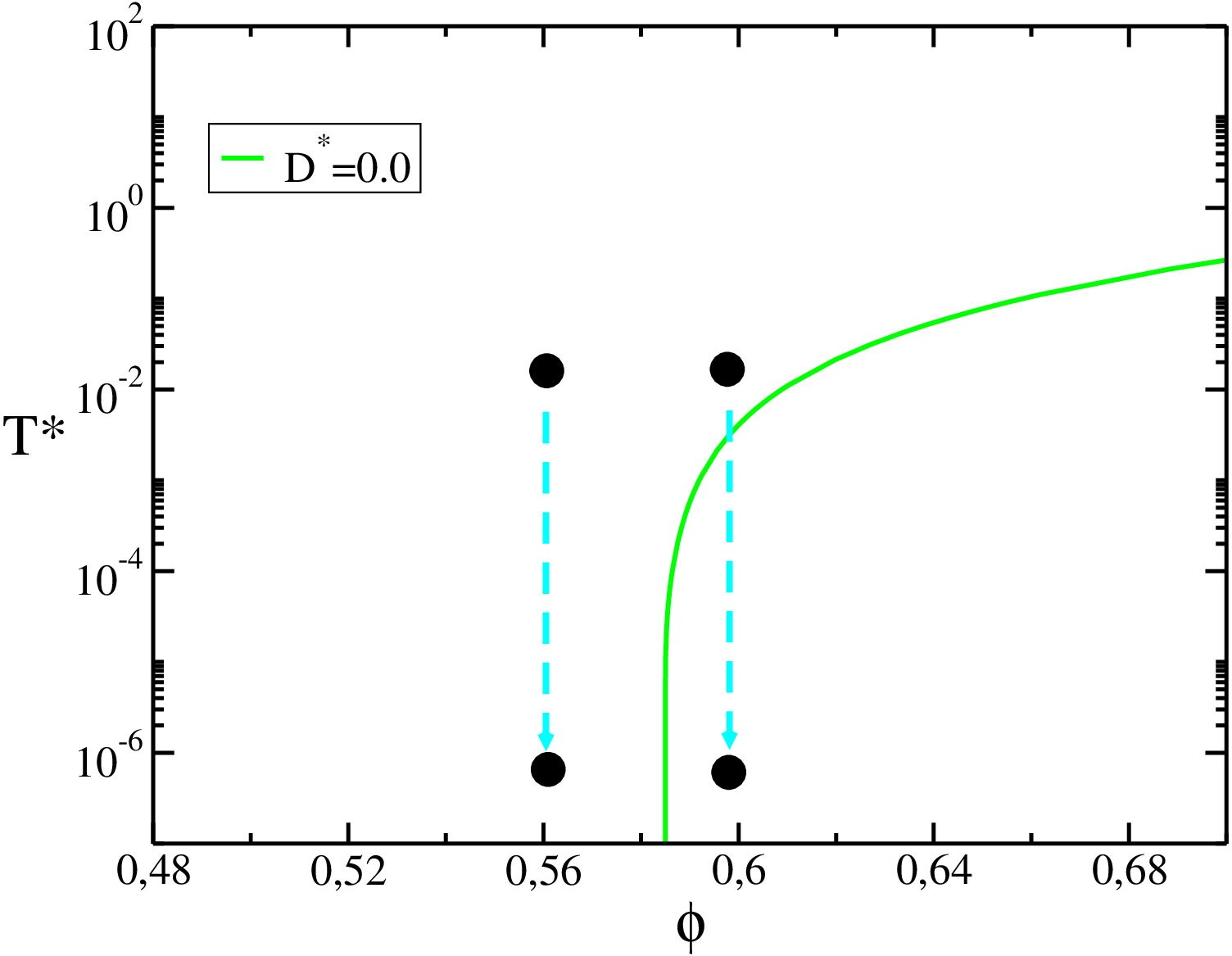}
\caption{Dynamic arrest line (or isodiffusivity curve with $D^*=0$) in the $(\phi,T^*)$ state space of the truncated Lennard-Jones fluid, Eq. (\ref{truncatedlj}), with $\nu=6$. The vertical downward arrows represent two fundamentally different classes of irreversible isochoric processes: in the first case the fixed volume fraction $\phi$ is smaller than the dynamic arrest volume fraction $\phi^{(a)}\ (=0.582)$, whereas in the second  $\phi$ is larger than $\phi^{(a)}$.} \label{fig1}
\end{figure}

We have employed in this manner the PY/VW approximation for the equilibrium static structure factor $S^{(eq)}(k;\overline{n},T)$ of the TLJ soft-sphere model, to  determine the region of its fluid-like ergodic states and the region of its dynamically arrested states. The resulting dynamic arrest transition line is represented by the solid curve in Fig. \ref{fig1} for the TLJ fluid with $\nu=6$, whose $T^* \to 0$ limit coincides with the dynamic arrest volume fraction $\phi^{(a)}$ of the hard sphere liquid, predicted to occur at $\phi^{(a)}=0.582$ \cite{gabriel}. As  indicated in the figure (and as explained at the end of the previous section), this transition line is also the iso-diffusivity curve corresponding to $D^*=0$.

\subsection{Method of solution of Ecs. (\ref{relsigmadif2pp}), (\ref{bdt})-(\ref{fluctsquench}) for equilibration.}\label{subsectionIII.3}

For concreteness, let us consider the case in which the system was initially prepared to be in the \emph{equilibrium} state corresponding to a point $(\phi,T^{*(i)})$ located in the fluid like region. We then have two fundamentally different possibilities, also illustrated in Fig. \ref{fig1}: either the final point $(\phi,T^{*(f)})$ lies in the ergodic region of the dynamic arrest diagram, or else, it lies in the region of dynamically arrested states. The first case is achieved, for example, if the volume fraction of the isochoric irreversible process is  smaller than the dynamic arrest volume fraction  $\phi^{(a)}=0.582$ of the hard sphere liquid. This isochoric quench $(\phi,T^{*(i)}) \to (\phi,T^{*(f)})$ will then eventually lead to the full equilibration of the system. In the second case, in which the fixed volume fraction  $\phi$ must be larger than $\phi^{(a)}=0.582$ (and the final temperature sufficiently low) the solution of Eqs. (\ref{relsigmadif2pp}), (\ref{bdt})-(\ref{fluctsquench}) will describe the irreversible aging of the glass-forming liquid quenched to a point inside the dynamically arrested region.

In either case, solving Ec. (\ref{relsigmadif2pp}) for $S(k;t)$ starts with the formal solution in Eq. (\ref{solsigmadkt}), written as
\begin{equation}
S^*(k;u)=S^{(i)}(k)e^{-\alpha (k)u}+ S^{eq}_f(k) \left(1-e^{-\alpha (k)
u}\right).  \label{solsdktexp}
\end{equation}
This expression interpolates $S^*(k;u)$ between its initial value $S^{(i)}(k) =S^{(eq)}(k;\phi,T^{*(i)})$
and its expected long-time equilibrium value $S^{eq}_f(k)\equiv S^{(eq)}(k;\phi,T^{*(f)})=
[\overline{n}^{(f)}\mathcal{E}^{(f)}(k)]^{-1}$. Clearly, the
solution $S(k;t)$ in Eq. (\ref{solsigmadkt}) can be written as
\begin{equation}
S(k;t)=S^*(k;u(t)),
\end{equation}
with $u(t)$ defined in Eq. (\ref{udt}). The inverse
function $t(u)$ is such that $u(t(u'))=u'$ and $t(u(t'))=t'$. The
differential form of Eq. (\ref{udt}) can be written as
$dt=du(t)/b(t)$. Upon integrating this equation, we have that
$t=\int_0^tdu(t')/b(t')$, which can also be written, after the
change of the integration variable $t'$, to $u'\equiv u(t')$, as
\begin{equation}
t(u) \equiv \int_0^u \frac{1}{b^*(u')}du',
\label{tdu00}
\end{equation}
with the function $b^*(u)$ defined as $b^*(u)=b(t(u))$. These general observations greatly simplify the mathematical analysis and the numerical method of solution of the
full NE-SCGLE theory under the particular conditions considered here.

To see this, let us consider a sequence $S^*(k;u_n)$ of snapshots of the static
structure factor, generated by the simple expression in Eq.
(\ref{solsdktexp}) when the parameter $u$ attains a sequence of equally-spaced
values $u_n$, say $u_n=n\Delta u$ (with a prescribed $\Delta u$ and
with $n=0,1,2,...$). The fact that $S(k;t)$ can be written as
$S(k;t)=S^*(k;u(t))$ implies that this sequence will be identical to the sequence $S(k;t_n)$
generated by the exact solution in Eq. (\ref{solsigmadkt}),
evaluated at a different sequence $t_n$ ($n=0,1,2,...)$, i.e., at a sequence of values of the time $t$, given by $t_n=\int_0^{u_n} [1/b^*(u')]du'$. In other words, the $n$th member of the sequence of static structure factors can be labeled either with the label $u_n$, as $S^*(k;u_n)$, or with the label $t_n$, as $S(k;t_n)$. For sufficiently small $\Delta u$, the discretized form of the previous relationship between $t_n$ and $u_n$ can be written as
\begin{equation}
t_{n+1}=t_n + \Delta u/ b^*(u_n). \label{recrel0}
\end{equation}
Thus, in practice what we do is to solve the self-consistent system of equations (\ref{bdt})-(\ref{lambdadk}) with $S(k;t)$ replaced by each snapshot $S(k;t_n)=S^*(k;u_n)$ of the sequence of static structure factors. This yields, among all the other dynamic properties, the sequence of values $b^*(u_n)$ of the function $b^*(u)$. This sequence can then be used in the recurrence relation in Eq. (\ref{recrel0}) to obtain the desired time sequence $t_n$, which allows us to ascribe a well-defined time label to the sequence $S(k;t_n)$ of static structure factors and to the sequence $b(t_n)$ of the instantaneous mobility $b(t)$. Of course, since the solution of equations (\ref{bdt})-(\ref{lambdadk}) yields all the dynamic properties, we also have in store the corresponding sequence of snapshots of dynamic properties such as $F(k,\tau;t_n)$, $F_S(k,\tau;t_n)$, the $\alpha$-relaxation time $\tau_\alpha(t_n)$, etc.

\section{Equilibration of soft-sphere liquids.}\label{sectionIV}

We have applied the protocol just described, which solves the full NE-SCGLE theory (Eqs. (\ref{solsigmadkt})-(\ref{lambdadk})), to the description of the isochoric irreversible evolution of the structure and the dynamics of the TLJ soft sphere liquid, after the instantaneous quench starting from an equilibrium fluid state. To continue the analysis, however, it is convenient to discuss separately the two mutually exclusive possibilities  illustrated by the two vertical arrows in Fig. \ref{fig1}. In this section we shall concentrate on the conceptually simplest case of the full equilibration of the system, and in the following section we shall discus the process of dynamic arrest.

\subsection{Ordered sequence of non-equilibrium static structure factors.}\label{subsectionIV.1}

\begin{figure}
\includegraphics[scale=.27]{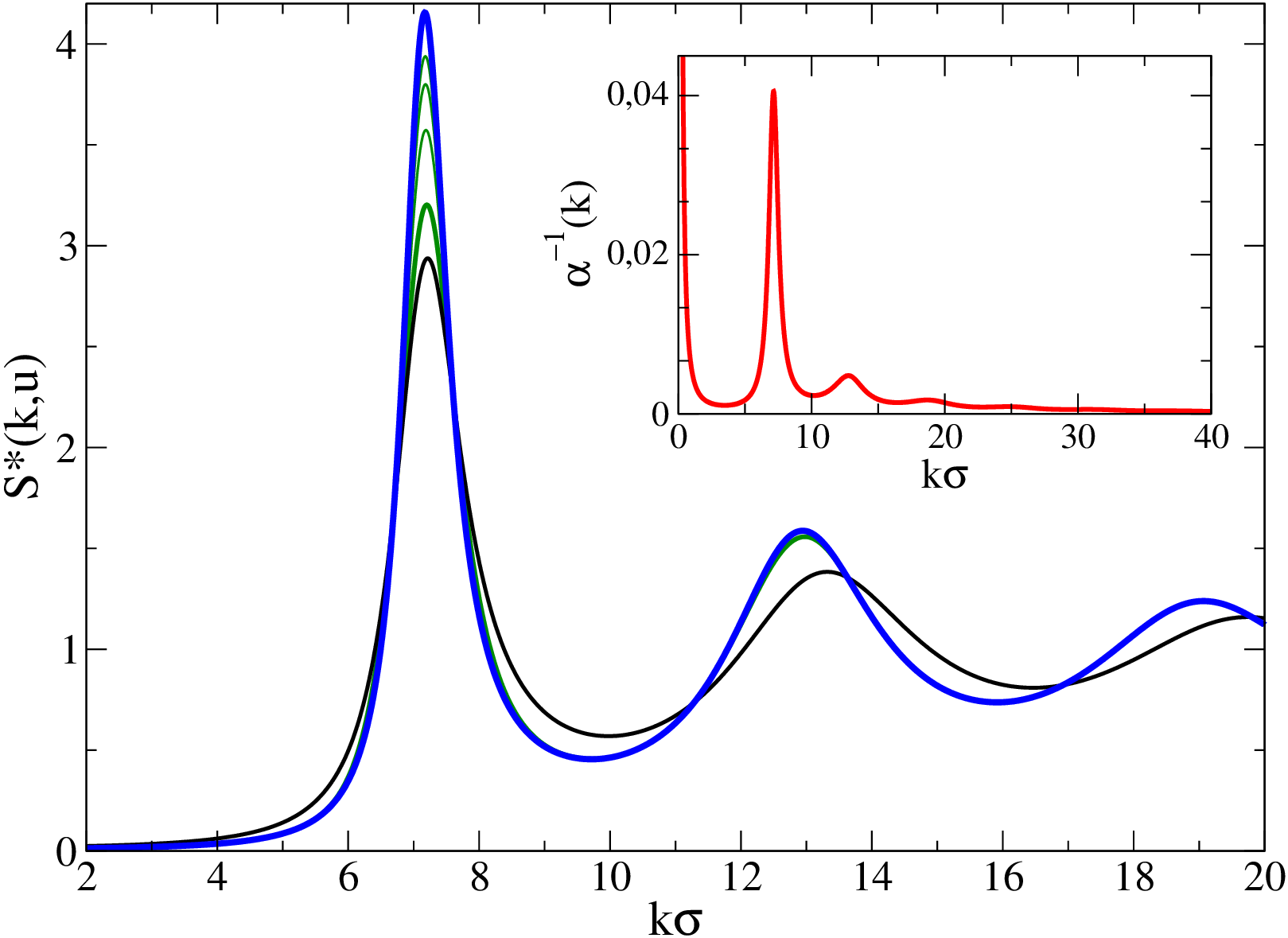}
\caption{Snapshots of the time-evolving static structure factor $S^*(k;u)$ corresponding to the quench $(\phi,T^{*(i)}) \to (\phi,T^{*(f)})$ at fixed volume fraction $\phi=0.56$ and $T^{*(i)}=0.1$, and $T^{*(f)} =0$. The darker thick (black) solid line is the initial structure factor $S^*(k;u=0)=S^{(i)}(k)$. The lighter thick (blue) solid line is the asymptotic limit $S^*(k;u\to\infty)=S^{eq}_f(k)$. The sequence of thinner (green) solid lines represent $S^*(k;u)$ for $u= \Delta u,\ 3\Delta u,\ 5\Delta u,$ and $7\Delta u$, with $\Delta u = 0.01$ (we use  $[\sigma^2/D^0]$  as the time unit). These  six structure factors also correspond, according to Eq. (\ref{tdu00}), to $S(k;t)$ for $t=0,\ 0.036,\ 0.1,\ 0.2, 0.533$, and $\infty$. Inset: Inverse relaxation constant $\alpha^{-1}(k)=S^{eq}_f(k)/2k^2D^0$. } \label{fig2}
\end{figure}

Let us thus illustrate the isochoric quench in which both, the initial and the final points, lie in the fluid-like region.  For concreteness, we consider a cooling process, $T^{(i)}>T^{(f)}$, such that  $b^{(i)} > b^{(f)}>0$, with  $T^{*(i)}=0.1$,  $\phi=0.56\ (< \phi^{(a)}=0.582)$, and with the final temperature corresponding to the deepest quench, $T^{*(f)} =0$. The initial and final equilibrium static structure factors, $S^{(i)}(k)=S^{(eq)}(k;\phi_1,T^{*(i)})$ and  $S^{eq}_f(k)=S^{(eq)}(k;\phi_1,T^{*(f)})$, are presented in Fig. \ref{fig2}. To visualize the transient non-equilibrium relaxation of $S(k;t)$, we generate a sequence of snapshots $S^*(k;u_n)$ using Eq. (\ref{solsdktexp}) with $u=u_n=n\Delta u$ ($n=0,1,2,...$) and with $\Delta u = 0.01[\sigma^2/D^0] \ (\approx 1/4\alpha (k)$, for $k=k_{max}$, the position of the main peak of $S^{eq}_f(k)$). From now on we shall use $[\sigma^2/D^0]$ as the time unit and $\sigma$ as the unit length. In Fig.  \ref{fig2} we include four representative intermediate snapshots of this sequence, corresponding to $u/\Delta u= 1,\ 3,\ 5,$ and 7. Let us emphasize that  although these snapshots of the transient structure factor are linear combinations of two equilibrium static structure factors (namely, $S^{(i)}(k)$ and  $S^{eq}_f(k)$), they themselves represent fully non-equilibrium structures.

Fig. \ref{fig2} exhibits the fact that within the resolution  $\Delta u = 0.01$  employed to visualize $S^*(k;u)$,  this non-equilibrium structure relaxes very quickly to its long-time equilibrium limit $S^{eq}_f(k)$ at most wave-vectors, except in two regions: in the vicinity of $k_{max}$, as appreciated in the figure,  and in the long-wavelength limit, $k \to 0$, not apparent in the main figure, but illustrated and discussed below. Thus, except in these two wave-vector domains, the non-equilibrium snapshots of $S^*(k;u)$ shown in the figure are already indistinguishable from $S^{eq}_f(k)$. The fact that for large wave-vectors, $k> k_{max}$, the structure $S^*(k;u)$ approaches very fast its final equilibrium value $S^{eq}_f(k)$ is understood by the fact that $\alpha(k)=2k^2D^0/S^{eq}_f(k)$ increases with $k^2$ while $S^{eq}_f(k)$ decreases from its maximum value towards its unit value at large $k$. To the left of $k_{max}$, on the other hand, although  $\alpha(k)$ decreases with $k^2$, there is a dramatic drop of the static structure factor from its large value at the main peak towards the very small value of $S^{eq}_f(k=0)$ of a strongly incompressible liquid. In support of this proposed scenario, in the inset of Fig.  \ref{fig2} we plot the inverse relaxation constant $\alpha^{-1}(k)=S^{eq}_f(k)/2k^2D^0$ as a function of $k$, which clearly exhibits a dominant peak at $k=k_{max}$, and a divergence at $k=0$. This explains the quick thermalization of $S^*(k;u)$ in both, the large wave-vector domain and in the moderately small wave-vector regime $0< k \lesssim k_{max}$.

\begin{figure}
\includegraphics[scale=.27]{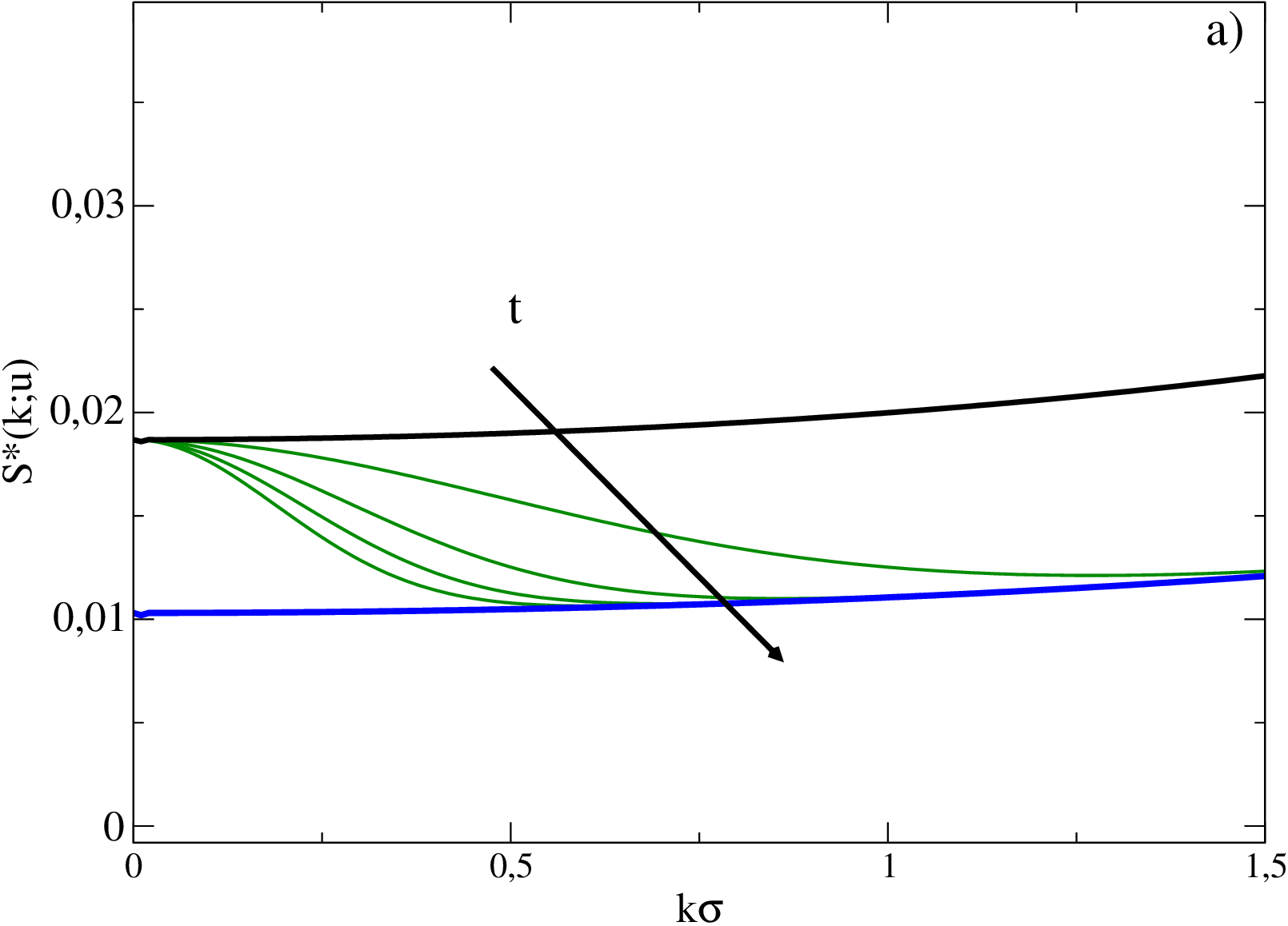}\hskip1.5cm
\includegraphics[scale=.27]{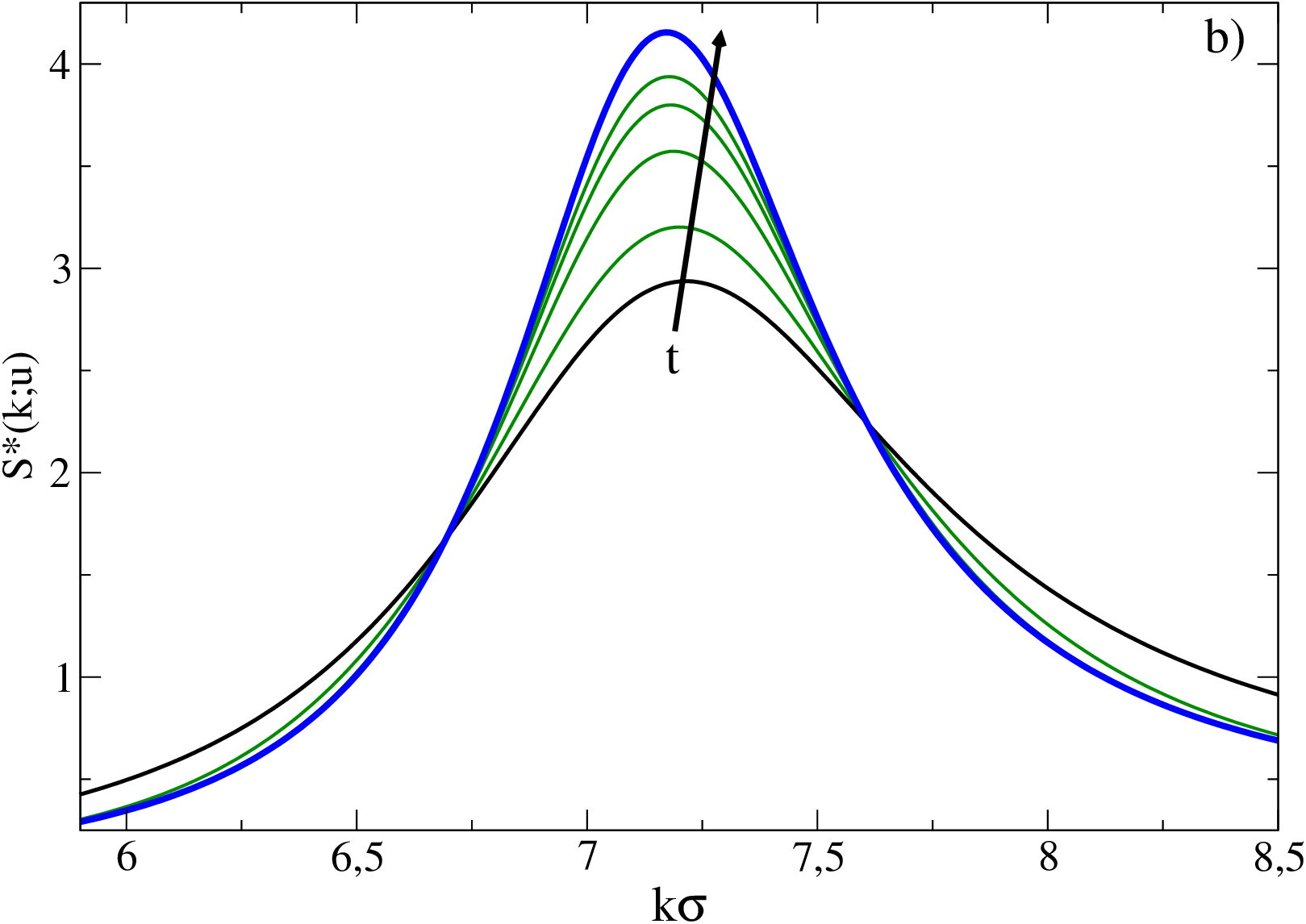}
\caption{Zoom of the snapshots $S^*(k;u_n)$ in Fig. \ref{fig2} corresponding to (a) the long-wavelength limit $k\to0$ and (b) to the neighborhood of the position $k_{max}$ of the main peak of $S^{eq}_f(k)$.} \label{fig3}
\end{figure}

In the really small wave-vector limit $k \to 0$, however, the $1/k^2$ divergence of  $u_{eq}(k)$ dominates, and prevents the thermalization of $S^*(k;u)$ within finite values of $u$. The crossover from this long-wavelength perfect slowdown, to the faster moderately-small wave-vector regime $k \lesssim k_{max}$, is revealed by zooming in at the small-$k$ behavior of the snapshots of $S^*(k;u)$, as illustrated in Fig.  \ref{fig3}(a). In contrast with this rather trivial long-wavelength slowing down, the slow relaxation at and around $k = k_{max}$ has its origin in the large value attained by $S^{eq}_f(k_{max})$, i.e., in the large strength of the interparticle correlations of spatial extent similar to the mean distance between the particles. Thus, this slowing down of the main peak of the $u$-evolving static structure factor is a non-equilibrium manifestation of the so-called cage effect.  In Fig.  \ref{fig3}(b) we present a zoom of the snapshots of $S^*(k;u)$ of Fig. \ref{fig2}, exhibiting in more detail the slower relaxation of the structure at these wave-vectors.

\subsection{Non-equilibrium $u$-dependence of $S^*(k;u)$ and $b^*(u)$.}\label{subsectionIV.2}

\begin{figure}
\includegraphics[scale=.27]{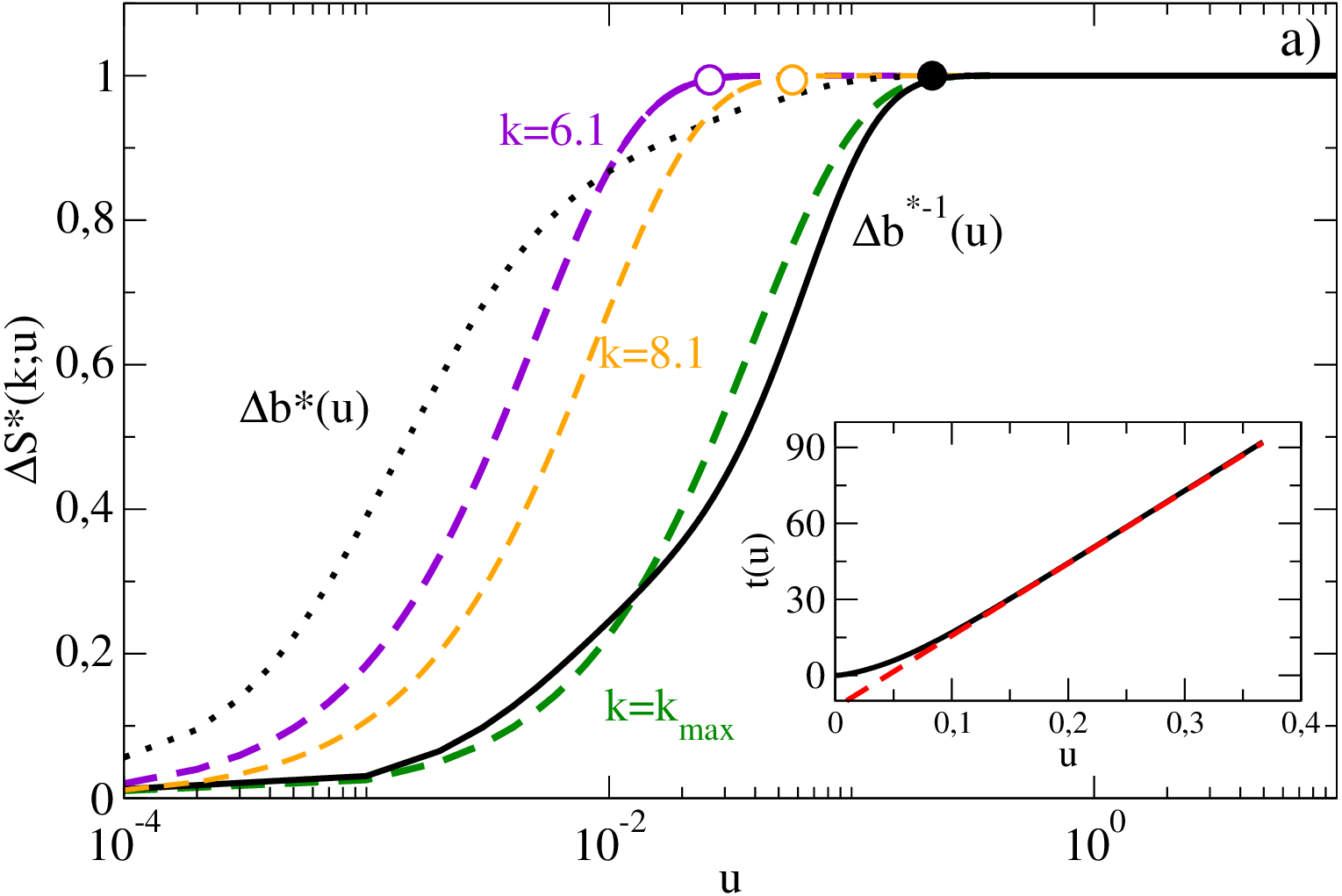}\hskip1cm
\includegraphics[scale=.27]{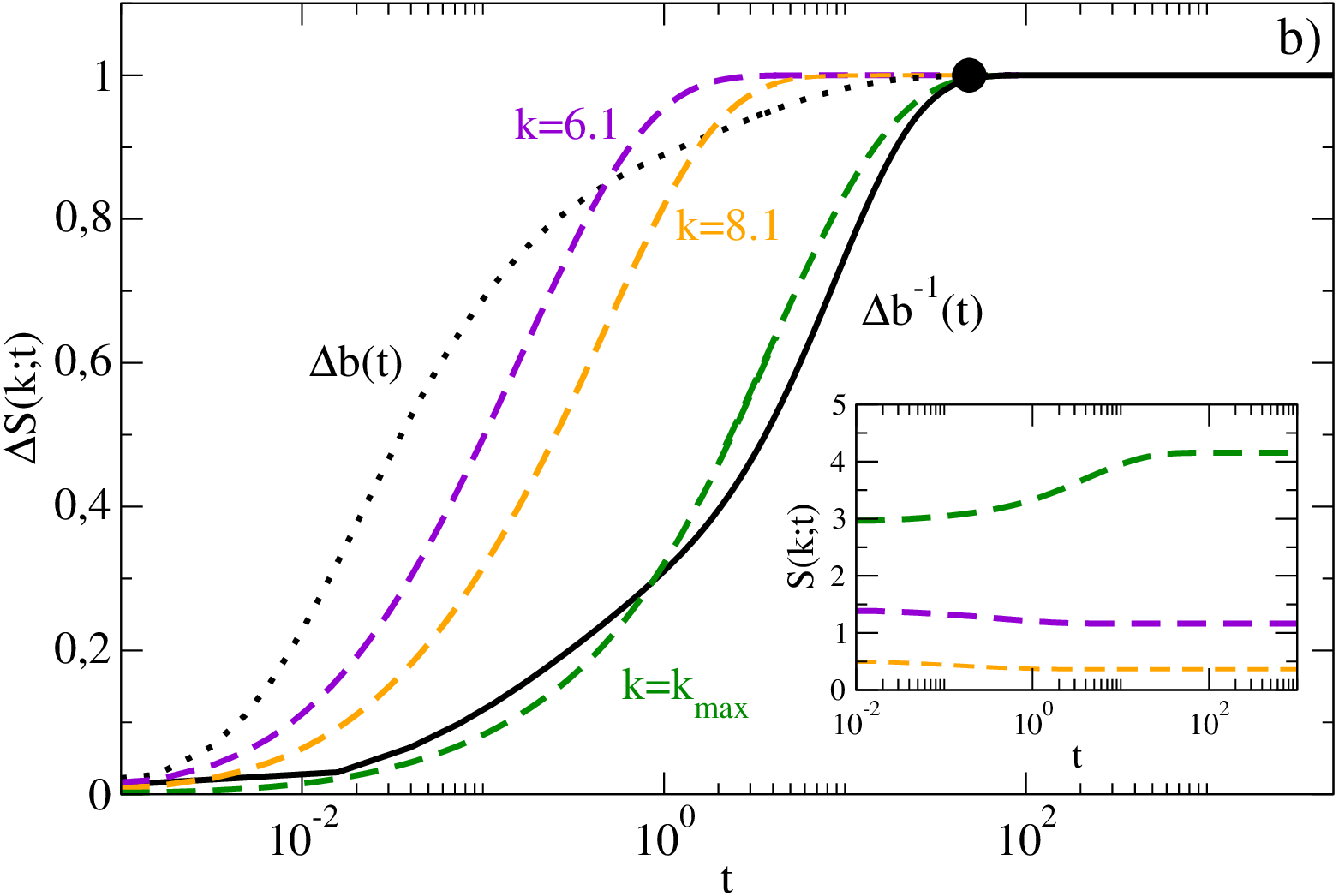}
\caption{(a) The function $\Delta S^*(k;u)=1-\exp [-\alpha (k)u]$ corresponding to three different wave-vectors, $k$ = 6.1, 7.26 $(=k_{max})$, and 8.1, plotted as a function of the parameter $u$ (dashed lines). The circles correspond to  $u=u_{eq}(k)=5\alpha^{-1}(k)$, where $[1-\Delta S^*(k;u_{eq}(k))]=e^{-5}$. The dotted and the solid dark lines represent, respectively, the functions $\Delta b^*(u)$ and  $\Delta b^{*-1}(u)$ (Ecs. (\ref{Deltabdu}) and  (\ref{Deltabinvdu})), and the solid circle corresponds to the estimated equilibration value $u^{eq}=u_{eq}(k_{max})$. The inset compares the function $t(u)$ calculated according to its exact definition in  Eq. (\ref{tdu00}) (solid line) and according to the approximation in Eq. (\ref{tduapprox}). (b) Same information as in (a), but now plotted as a function of the actual \emph{evolution} time $t\equiv \int_0^u \frac{1}{b^*(u')}du'$. In the main figure we plot $\Delta S(k;t)$ and in the inset we plot $ S(k;t)$.} \label{fig3p}
\end{figure}

Let us notice that the simple expression for $S^*(k;u)$ in Eq. (\ref{solsdktexp}), which interpolates this function of $u$ between $S^{(i)}(k)$ and $S^{eq}_f(k)$, may be written as
\begin{equation}
\Delta S^*(k;u)\equiv \frac{S^*(k;u)-S^{(i)}(k)}{S^{eq}_f(k)-S^{(i)}(k)}=1-e^{-\alpha (k)u}. \label{solsdktexp2}
\end{equation}
This means that if we plot the static structure factor $S^*(k;u)$ as $\Delta S^*(k;u)$ vs. the $k$-dependent variable $[\alpha(k) u]$, the results for all the wave-vectors $k$ must collapse onto a master curve independent of $k$ and of the initial and final values $S^{(i)}(k)$ and $S^{eq}_f(k)$. In fact, such a master curve will be essentially a simple exponential function. This simplicity, however, will be partially lost if we plot $\Delta S^*(k;u)$ directly as a function of the parameter $u$, since such exponential function, $e^{-\alpha (k)u}$, will decay with $u$ at a different rate  $\alpha (k)$ for different values of the wave-vector $k$, as illustrated in Fig. \ref{fig3p}(a). In fact, if we define a  $k$-dependent\emph{ equilibration} value $u_{eq}(k)$ by the condition $e^{-\alpha (k)u_{eq}(k)} \approx e^{-5}$, we have that $u_{eq}(k)\equiv 5\alpha^{-1}(k)$. Thus, except for the arbitrary factor of 5, the inset of Fig. \ref{fig2}(a) exhibits the wave-vector dependence of $u_{eq}(k)$. There we see that  $u_{eq}(k)$ attains its largest value at the wave-vector $k_{max}$, corresponding to the position of the main peak of $S^{eq}_f(k)$. This  slowest mode imposes the pace of the overall equilibration process, thus characterized by the $k$-independent \emph{equilibration } value $u^{eq}\equiv u_{eq}(k_{max})=5S^{eq}_f(k_{max})/2k_{max}^2D^0$.

The previous discussion illustrates the properties of the ordered sequence $S^*(k;u_n)$ of snapshots of the function $S^*(k;u)$ for equally-spaced values $u_n$ of the parameter $u$.
This sequence of snapshots, however, do not fully reveal the most important features of the real relaxation scenario implied by the solution (\ref{solsigmadkt}) of Eq. (\ref{relsigmadif2pp}), which provides $S(k;t)$ as a function of the actual evolution time $t$. Nevertheless, since $S(k;t_n)=S^*(k;u(t_n))$, these features are fully revealed by simply relabeling the referred sequence $S^*(k;u_n)$ using the (not equally-spaced) sequence of labels $t_n$ given by  the recurrence relation in Eq. (\ref{recrel0}). This results in the sequence $S(k;t_n)$ of snapshots that describes the actual time evolution of $S(k;t)$. In order to carry out this program, however, we must first determine the sequence $b^*(u_n)$ needed in the referred recurrence relation.

As indicated before, from any sequence of snapshots $S^*(k;u_n)$, with $u=n\Delta u$ ($n=0,1,2,...$), we may generate a sequence $b^*(u_n)$ of values of the time-dependent mobility $b^*(u)$ by  solving the self-consistent system of equations (\ref{bdt})-(\ref{lambdadk}) with $S(k;t)$ replaced by $S^*(k;u_n)$ for each snapshot. The resulting sequence $b^*(u_n)$ is a discrete representation of the function $b^*(u)$, shown in Fig. \ref{fig3p}(a), whose resolution in  the parameter $u$ may be improved arbitrarily by taking $\Delta u$ as small as needed. The first feature to notice in the result of this procedure is the fact that $b^*(u)$ decays monotonically from its initial value $ b_i$ to its final value  $b_f>0$. This implies  that the system will always remain fluid-like and will have no impediment to reach its expected equilibrium state. We also find that the function $b^*(u)$ attains its asymptotic value $b_f$ for $u> u^{eq}\ (\approx 0.2$  for the quench illustrated in the figure).

In  Fig. \ref{fig3p}(a) we also present the results for $b^*(u)$  plotted as
\begin{equation}
\Delta b^*(u)\ \equiv \frac{b^*(u)-b_i}{b_f-b_i}. \label{Deltabdu}
\end{equation}
We see that this plot does not exhibit any simple relationship between the decay of $\Delta b^*(u)$ and the decay of $\Delta S^*(k;u)$. In the same figure, however, the results for $b^*(u)$ are plotted as
\begin{equation}
\Delta b^{*-1}(u)\ \equiv \frac{b^{*-1}(u)-b^{-1}_i}{b^{-1}_f-b^{-1}_i}. \label{Deltabinvdu}
\end{equation}
Plotted in this manner we observe a more apparent correlation between the decay of both, $\Delta S^*(k_{max};u)$ and $\Delta b^{*-1}(u)$, with the parameter $u$. This feature remains, of course, when these properties are expressed as functions of the actual evolution time $t$, as we now see.

\subsection{Real-time dependence of $S(k;t)$ and $b(t)$.}\label{subsectionIV.3}

Once we have determined the function $b^*(u)$, using the expression for $t(u)$ in  Eq. (\ref{tdu00}), or its discretized version in the recursion relation of  Eq. (\ref{recrel0}), we can determine the desired real-time evolution of $S(k;t)$ and $b(t)$. In this manner we determine that in our illustrative example the sequence $u=0, 0.01,\ 0.03,\ 0.05,$ and $0.07$ corresponds to the sequence $t=\ 0,\ 0.036,\ 0.1,\ 0.2,$ and 0.533. In the inset of Fig.  \ref{fig3p}(b) we present the resulting time-evolution of $S(k;t)$ for the same three wave-vectors as in  Fig.  \ref{fig3p}(a). This inset emphasizes the fact that $S(k;t)$ evolves monotonically from its initial value $ S^{(i)}(k)$ to its final value  $S^{eq}_f(k)$, sometimes increasing and sometimes decreasing,  depending on the wave-vector considered. In order to exhibit a less detail-dependent scenario, in the main frame of Fig.  \ref{fig3p}(b) we present the same information, but formatted as $\Delta S(k;t)$, which is the  re-labeled version ($u \to t=t(u)$) of $\Delta S^*(k;u)$ in Eq. (\ref{solsdktexp2}), namely, as
\begin{equation}
\Delta S(k;t)\equiv \frac{S(k;t)-S^{(i)}(k)}{S^{eq}_f(k)-S^{(i)}(k)}. \label{solsdktexp3}
\end{equation}
We similarly relabel the definitions of $\Delta b^*(u)$ and $\Delta b^{*-1}(u)$ in Eqs. (\ref{Deltabdu}) and  (\ref{Deltabinvdu}) to define the functions $\Delta b(t)$ and $\Delta b^{-1}(t)$, which are also plotted in Fig.  \ref{fig3p}(b). The comparison of this figure with Fig.  \ref{fig3p}(a) indicates that, except for the stretched metric of $t$, the overall scenario described by the $u$-dependence illustrated in Fig.  \ref{fig3p}(a) is preserved in the $t$-dependence illustrated in Fig.  \ref{fig3p}(b).

Let us notice in particular that the existence of the equilibration value $u^{eq}$ of the parameter $u$, beyond which $b^*(u)\approx b_f$, allows us to define an \emph{equilibration} time, $t^{eq}$, as the time that corresponds to $u^{eq}$ through Eq. (\ref{tdu00}),
\begin{equation}
t^{eq} \equiv \int_0^{u^{eq}}\frac{du'}{b^*(u')}. \label{equiltime1}
\end{equation}
For our specific illustrative example, this yields $t^{eq}\approx48$. The fact that $b^*(u)\approx b_f$ for $u \gtrsim u^{eq}$ implies, according to Eq. (\ref{tdu00}), that for $u>u^{eq}$ the function $t(u)$ will be linear in $u$, i.e.,
\begin{equation}
t(u) \approx -a(u^{eq}) + b_f^{-1}u, \label{tduapprox}
\end{equation}
with $a(u^{eq})\equiv \int_0^{u^{eq}}[1/b_f-1/b^*(u')]du'$. In the inset of Fig.  \ref{fig3p}(a) we compare this asymptotic expression, applied to our illustrative case (for which $b_f^{-1}=285$ and $a(u^{eq})=12.8$), with the actual $t(u)$ calculated from Eq. (\ref{tdu00}).

\subsection{Irreversibly-evolving dynamics.}\label{subsectionIV.4}

\begin{figure}
\includegraphics[scale=.27]{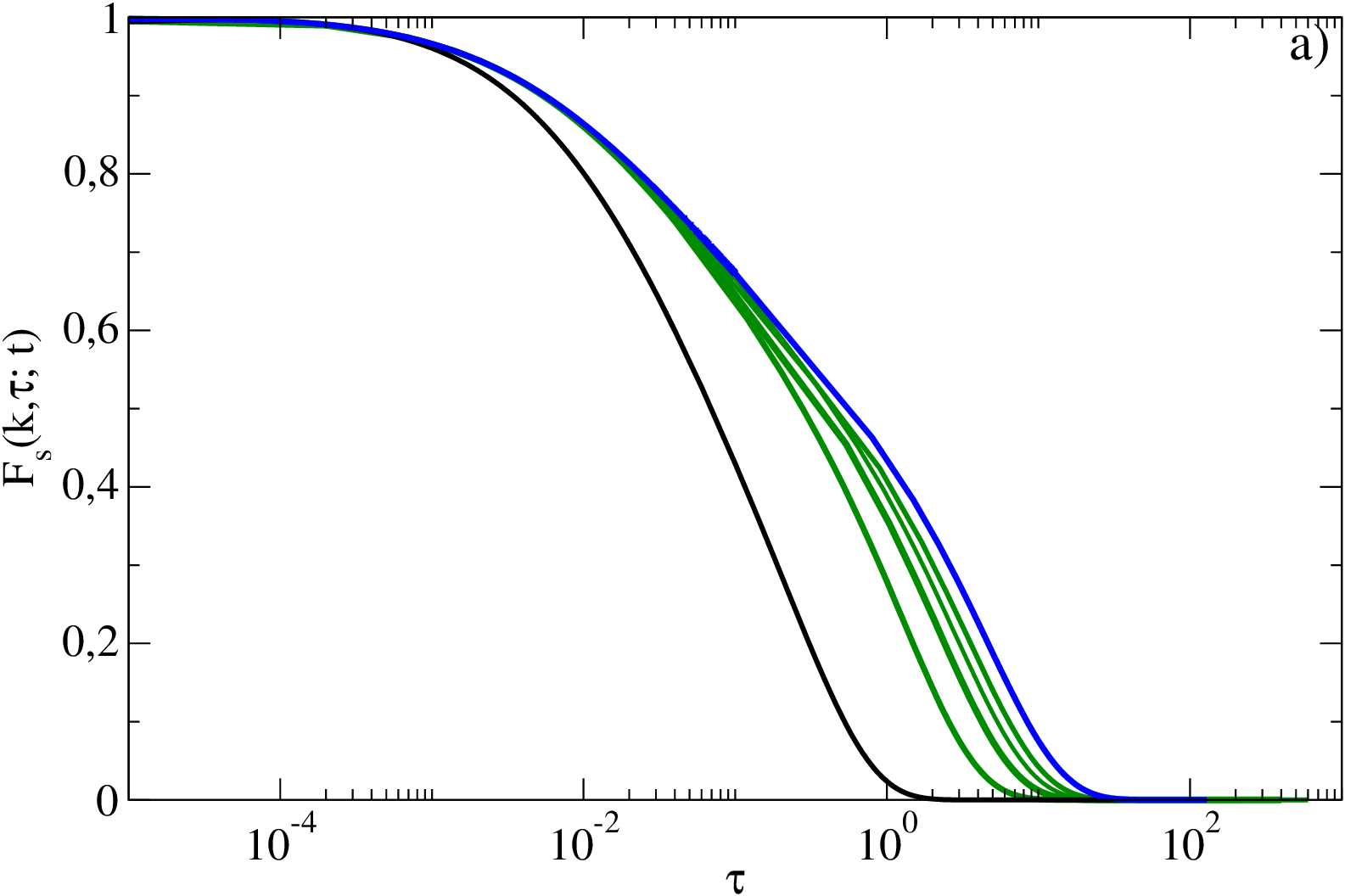}
\includegraphics[scale=.27]{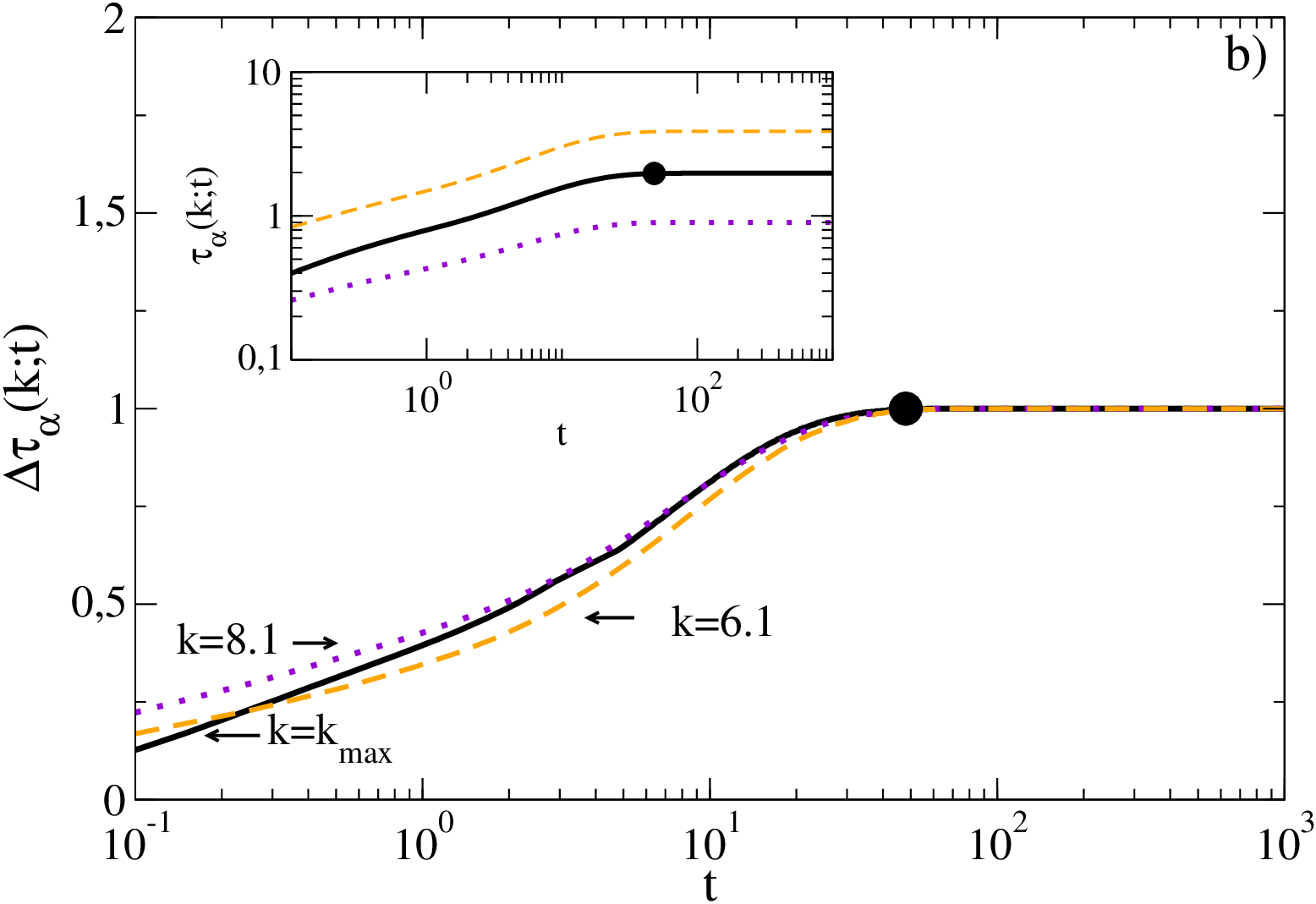}
\caption{(a) Snapshots of the non-equilibrium self intermediate scattering function $F_S(k,\tau;t)$ at $k=k_{max}$, corresponding to the equilibration process in Fig. \ref{fig2} for evolution times $t=0,\ 0.036,\ 0.1,\ 0.2, 0.533$ and $\infty$.  (b) Non-equilibrium evolution of the dimensionless $\alpha$-relaxation time, displayed as  $\tau_{\alpha}(k;t)$ itself (inset) and formatted as in Eq. \ref{taualphastar} (main figure). The solid circle represents the equilibration point ($t^{eq},\tau_\alpha(k_{max};t^{eq})$). } \label{fig5}
\end{figure}

Since for each snapshot of the static structure factor $S(k;t)$ the solution of Eqs. (\ref{bdt})-(\ref{lambdadk}) determines a snapshot of each of the dynamic properties of the system, the process of equilibration may also be observed, for example, in the $t$-evolution of the collective and self intermediate scattering functions, $F(k,\tau;t)$ and $F_S(k,\tau;t)$. In  Fig.  \ref{fig5}(a) we illustrate this irreversible time-evolution with the snapshots of the self-ISF $F_S(k_{max},\tau;t)$, corresponding to the same set of evolution times $t_n$ as the snapshots of $S(k;t)$ in Fig. \ref{fig2}. We see that the function $F_S(k,\tau;t)$ starts from its initial value $F_S(k,\tau;t=0)=F_S^{eq}(k,\tau;\phi,T_i=0.1)$, and quickly evolves with waiting time $t$ towards the vicinity of its final equilibrium value $F_S(k,\tau;t=\infty)=F_S^{eq}(k,\tau;\phi,T_f=0)$.

The equilibration process of $F_S(k,\tau_{\alpha};t)$ can be best summarized in terms of the dependence of the $\alpha$-relaxation time $\tau_{\alpha}(k;t)$ as a function of the evolution time $t$. The $\alpha$-relaxation time may be defined by the condition
\begin{equation}
F_S(k,\tau_{\alpha};t) = 1/e.
\label{taualphadef}
\end{equation}
The dependence of $\tau_{\alpha}(k;t)$ on the evolution time $t$ can be extracted from a sequence of snapshots of $F_S(k,\tau_{\alpha};t)$, such as those in  Fig. \ref{fig5}(a). The results are illustrated in  Fig.  \ref{fig5}(b), in which the solid line corresponds to $\tau_{\alpha}(k_{max};t)$. The solid circle indicates the crossover from the $t$-regime where  $\tau_{\alpha}(k_{max};t)$ is still in the process of equilibration, to the regime where it has reached its final equilibrium value  $\tau_{\alpha}^{(f)}(k_{max})\equiv \tau_{\alpha}^{eq}(k_{max};\phi,T_f)$. In the inset of the figure we plot $\tau_{\alpha}(k_{max};t)$ itself and in the main figure we plot the same information, but
formatted as
\begin{equation}
\Delta \tau_{\alpha}(k;t)\ \equiv \frac{\tau_{\alpha}(k;t)-\tau_{\alpha}^{(i)}(k)}{\tau_{\alpha}^{(f)}(k)-\tau_{\alpha}^{(i)}(k)}, \label{taualphastar}
\end{equation}
with  $\tau_{\alpha}^{(i)}(k)\equiv \tau_{\alpha}^{eq}(k;\phi,T_i)$.
In the same figure we also exhibit similar results corresponding to two additional wave-vectors, different from $k_{max}$ (dashed lines). These results show that the equilibration time of  $\tau_{\alpha}(k;t)$ for these three wave-vectors is  largely independent of $k$, and can be well approximated by the equilibration time $t^{eq}$ defined in Eq. (\ref{equiltime1}), in contrast with the notorious wave-vector dependence of the predicted evolution of the static structure factor illustrated in Fig. \ref{fig3p}(b).

\begin{figure}
\includegraphics[scale=.27]{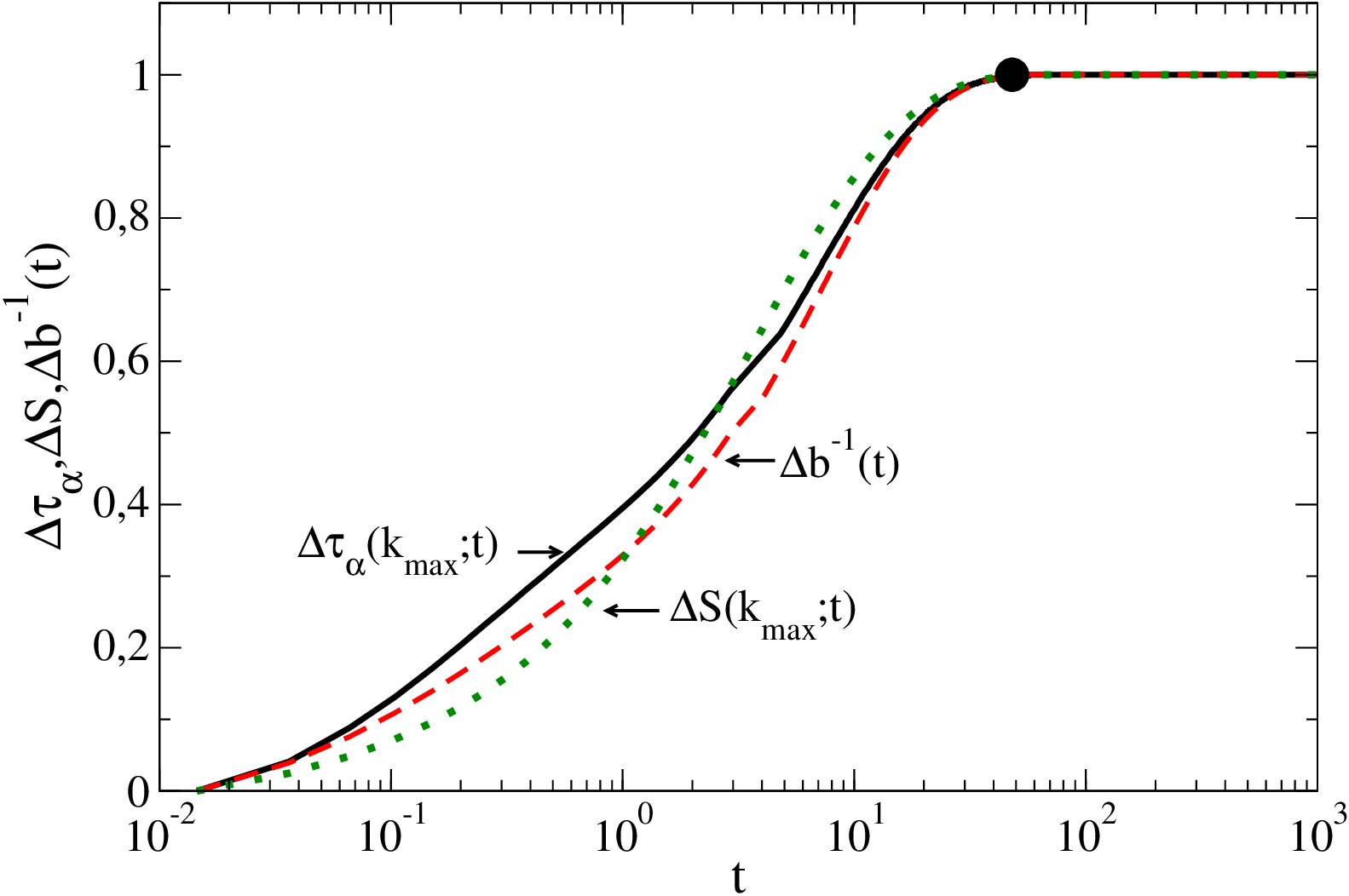}
\caption{Time-evolving $\alpha$-relaxation time
$\tau_{\alpha}(k_{max};t)$ (solid line) mobility $b(t)$ (dash
line) and static structure factor $S(k_{max};t)$ (dotted line) as
a function of evolution time $t$, and expressed as $\Delta
\tau_{\alpha}(k_{max};t)$,  $\Delta b^{-1}(t)$, and $\Delta
S(k_{max};t)$. } \label{fig6}
\end{figure}

Let us finally mention another theoretical prediction regarding the kinetics of the equilibration process.  This refers to the similarity of the equilibration kinetics exhibited by  the time-dependent mobility $b(t)$, the $\alpha$-relaxation time $\tau_{\alpha}(k;t)$ at all wave-vectors, and the static structure factor $S(k_{max};t)$ at the wave-vector $k_{max}$, when plotted in terms of the reduced properties  $\Delta b(t)$, $\Delta \tau_{\alpha}(k;t)$, and $\Delta S(k_{max};t)$. This similarity is exhibited in Fig.  \ref{fig6}  for the illustrative quench at fixed $\phi=0.56$, and means that indeed the evolution of $S(k_{max};t)$ (which is  slower than the evolution of $S(k;t)$ for other wave-vectors) sets the overall relaxation rate exhibited by the dynamic properties $\Delta b(t)$ and $\Delta \tau_{\alpha}(k;t)$. Thus, from this point of view, we may use either of these characteristic dynamic properties to describe the predicted kinetics of the equilibration process.

\subsection{Dependence on the initial temperature of the quench.}\label{subsectionIV.5}

Up to this point we have illustrated the main features of the isochoric quench $(\phi,T^{*(i)}) \to (\phi,T^{*(f)})$ at fixed volume fraction $\phi=0.56$, using for concreteness the values $T^{*(i)}=0.1$ and $T^{*(f)}=0$. We are now ready to analyze how the scenario just described depends on the initial temperature $T^{*(i)}$ and on the volume fraction $\phi$ at which we perform the quench. Let us start by considering the dependence on $T^{*(i)}$. Rather than attempting a comprehensive illustration of this dependence in terms of the evolution of the static structure factor $S(k;t)$ and of the various dynamic properties, we use the dimensionless mobility $b(t)$ as a representative property bearing the essential information about the equilibration process. This $k$-independent property determines the mapping from the parameter $u$ to the real time $t$, through the definition of the functions $u(t)$ and $t(u)$ in Eqs. (\ref{udt}) and (\ref{tdu00}).

\begin{figure}
\includegraphics[scale=.27]{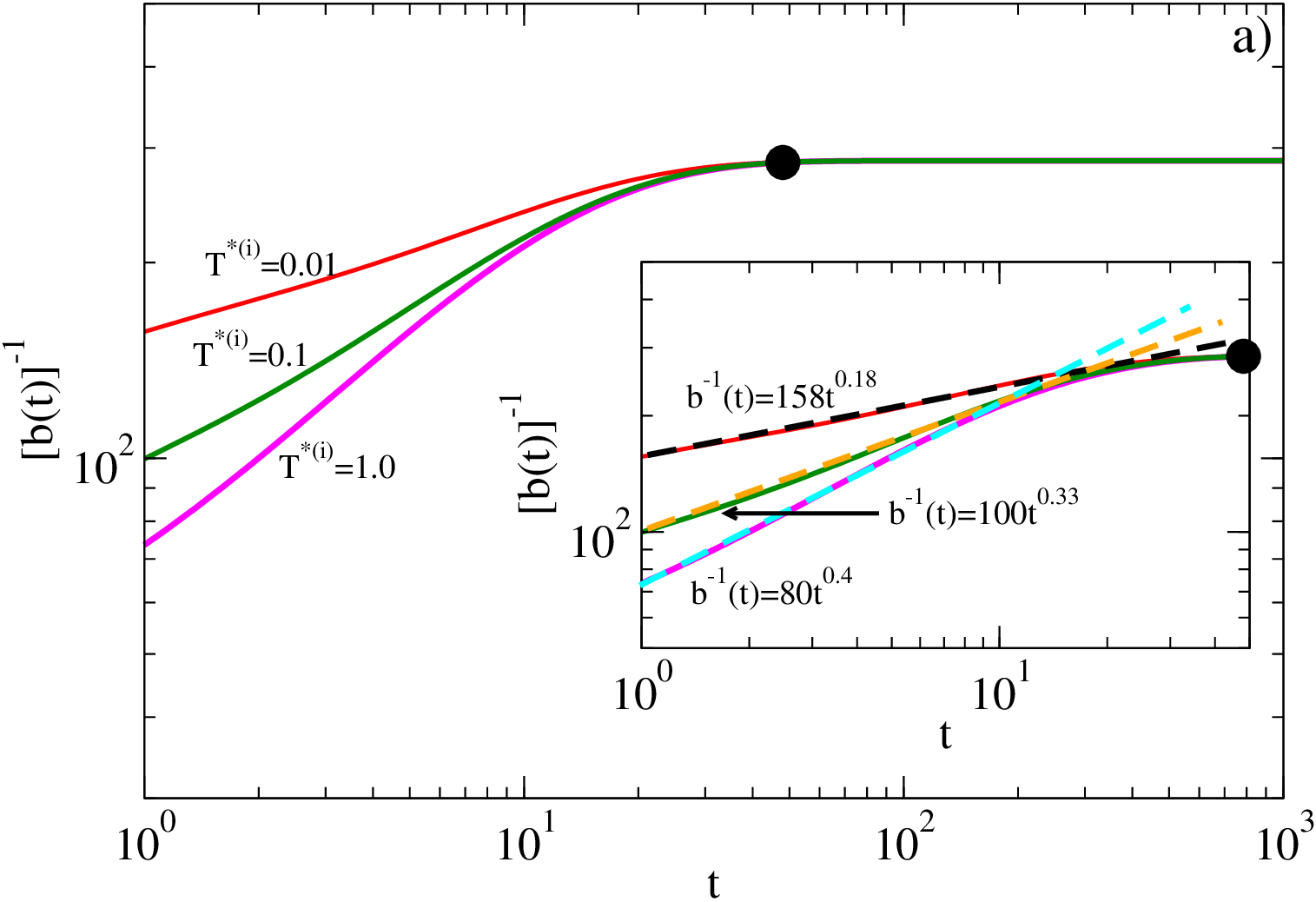}
\includegraphics[scale=.27]{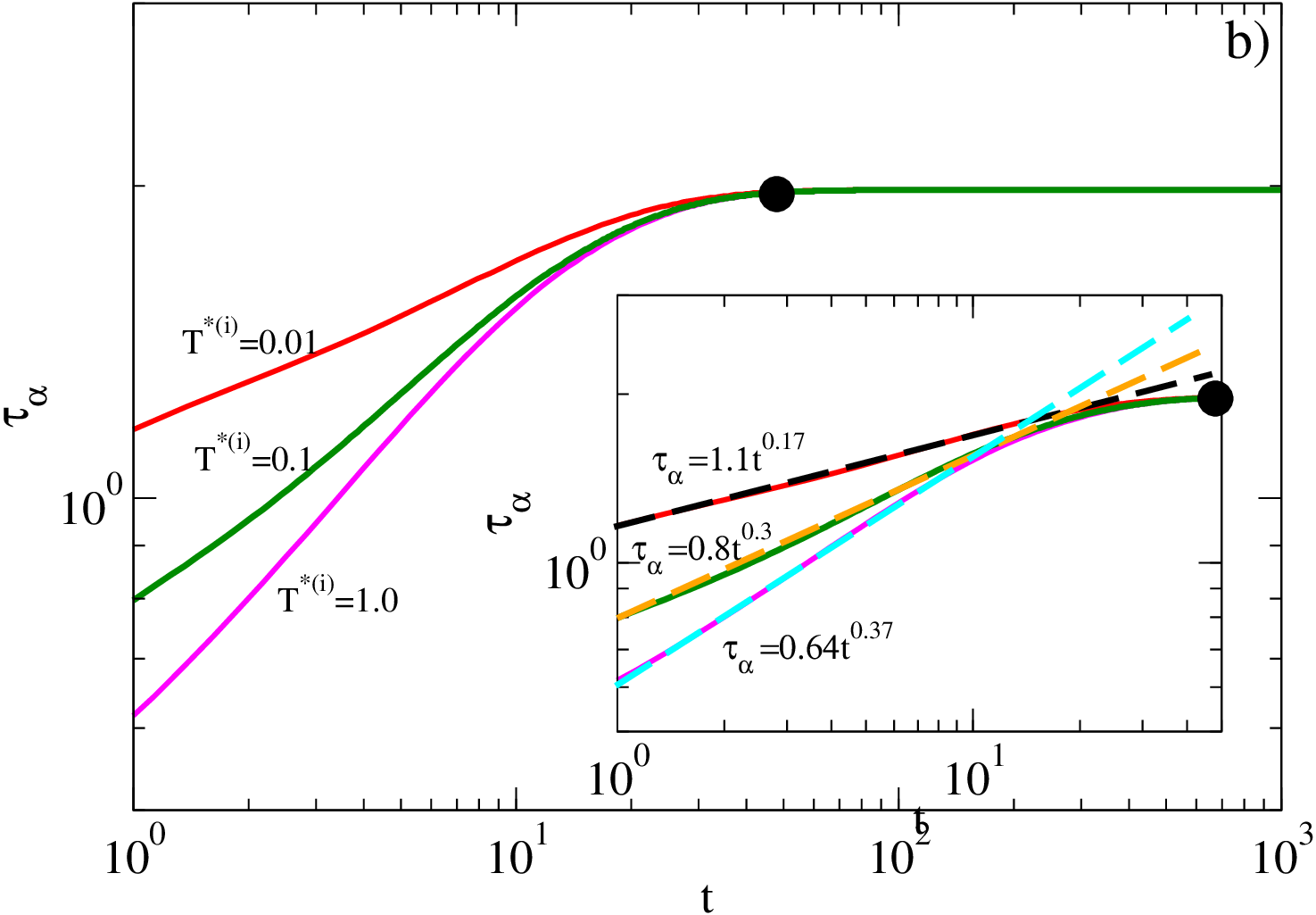}
\caption{ Non-equilibrium, time-dependent (a) mobility $b(t)$ and
(b) $\alpha$-relaxation time $\tau_\alpha(t)$, as a function of
evolution time $t$,  for the isochoric quench $(\phi,T^{*(i)}) \to
(\phi,T^{*(f)}=0)$ at fixed volume fraction $\phi=0.56$, for three
different initial temperatures, $T^{*(i)}$= 1.0, 0.1, and 0.01.
The insets exhibit a power law fit of the transient before
saturation.} \label{fig7}
\end{figure}

Thus, in Fig. \ref{fig7}(a) we present plots of  $b^{-1}(t)$ as a function of $t$ for three representative values of the initial temperature $T^{*(i)}$, namely, $T^{*(i)}$= 1.0, 0.1, and 0.01, keeping the same final temperature $T^{*(f)}=0$ and the same volume fraction $\phi$=0.56. This figure reveals two remarkable features. In the first place, the equilibration time $t^{eq}$ seems to be rather insensitive to the temperature $T^{*(i)}$ of the initial state. In other words, the system will reach the final equilibrium state in about the same time, $t^{eq} \approx 48$, no matter if the initial temperature is $T^{*(i)}$= 1.0, 0.1, or 0.01. To emphasize this feature we have highlighted the common equilibration point of the three curves. The second remarkable feature is that during the transient stage of the equilibration process, the evolution of $b^{-1}(t)$ as a function of $t$ follows approximately a power law, $b^{-1}(t)\approx At^{x}$, with the exponent $x$ and the amplitude  $A$ depending on the initial temperature  $T^{*(i)}$. In the inset of the figure we exhibit the power law fit of the transient, indicating the resulting value of the exponent $x$ and amplitude $A$.

Exactly the same trend is also reflected in the evolution of the intermediate scattering function, as observed in the results for  the $\alpha$-relaxation time shown in Fig. \ref{fig7}(b). This information is important, since many times it is this dynamic parameter what is monitored in simulations and in some experiments.

\subsection{Dependence on the volume fraction of the quench.}\label{subsectionIV.6}

\begin{figure}
\includegraphics[scale=.27]{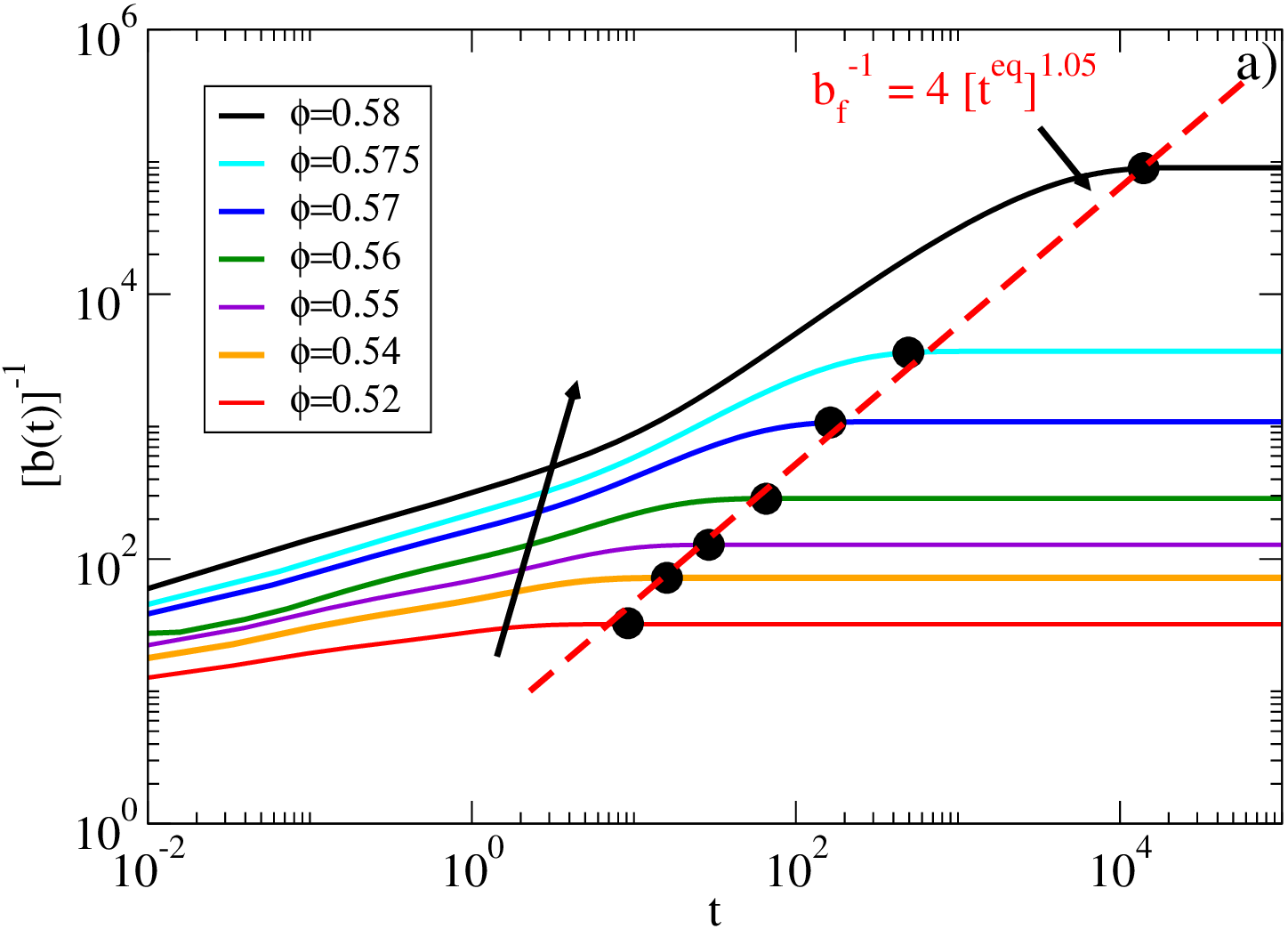}\hskip1cm
\includegraphics[scale=.27]{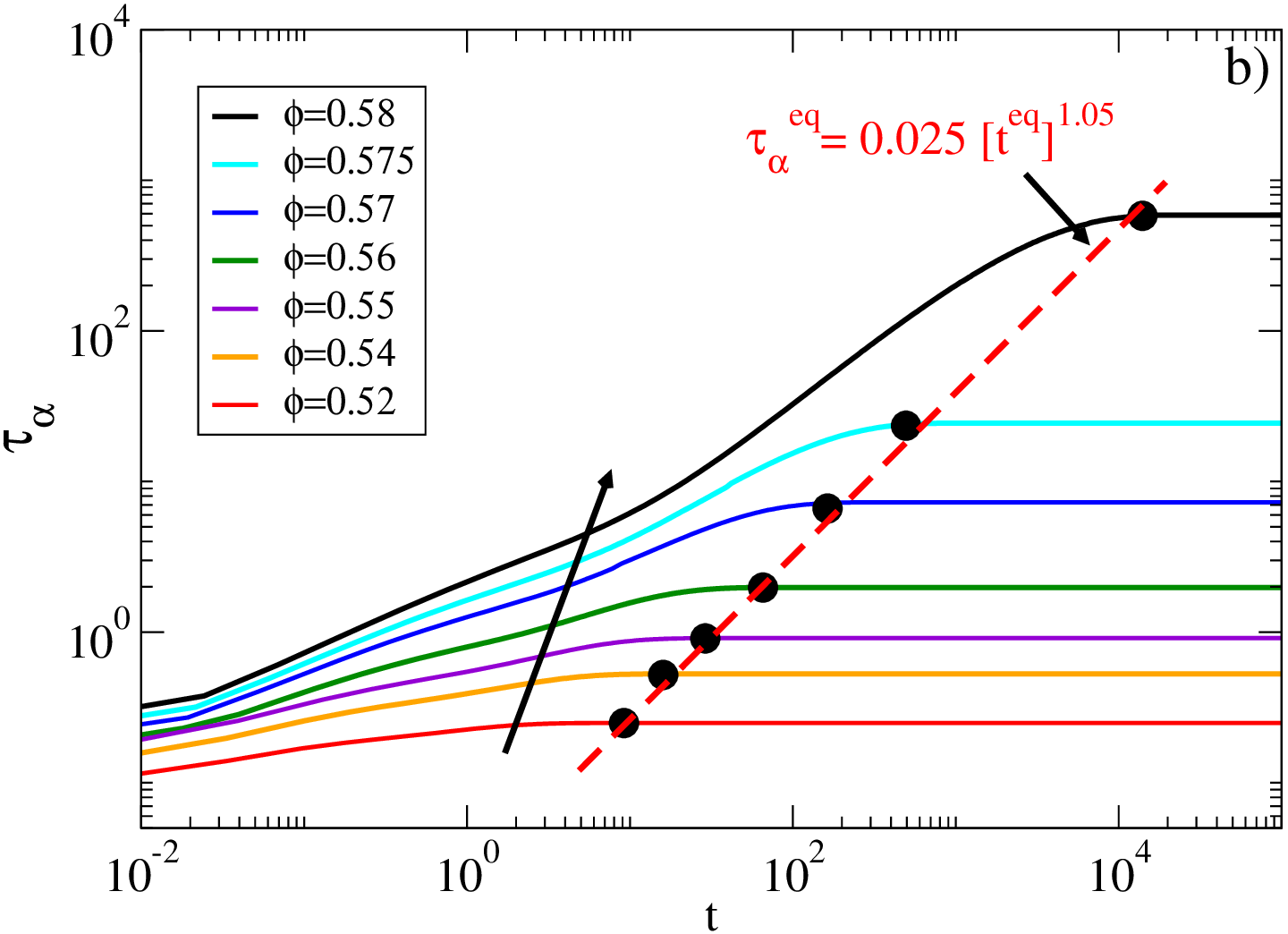}
\caption{Non-equilibrium, time-dependent (a) mobility $b^{-1}(t)$
and (b) $\alpha$-relaxation time $\tau_\alpha(t)$, as a function
of evolution time $t$ for various volume fractions corresponding
to the metastable regime of the hard-sphere liquid. The arrows indicate the direction of increasing $\phi$. The dashed lines are the power laws $ b_f^{-1}(\phi) \approx 4 \times [t^{eq}(\phi)]^{1.05}$ and $ \tau^{eq}_{\alpha}(\phi) \approx 0.025 \times [t^{eq}(\phi)]^{1.05}$.} \label{fig8}
\end{figure}

Let us now discuss the dependence of the equilibration process on the volume fraction $\phi$. Once again we first use the time evolution of $b(t)$ to illustrate this dependence. In Fig. \ref{fig8}(a) we plot  $b^{-1}(t)$ as a function of $t$ for a set of values of the volume fraction $\phi$, corresponding to the metastable regime of the hard-sphere liquid. According to these results, the inverse mobility $b^{-1}(t;\phi)$ reaches its equilibrium value $b_f^{-1}(\phi)$ after a $\phi$-dependent equilibration time $t^{eq}(\phi)$. To emphasize this prediction, the solid circles in the figure highlight the points $(t^{eq}(\phi), b_f^{-1}(\phi))$. These highlighted points, as indicated in the figure, align themselves to a good approximation along the dashed line of the figure, corresponding to the approximate relationship $ b_f^{-1}(\phi) \approx 4 \times [t^{eq}(\phi)]^{1.05}$

This relationship between the equilibration time $t^{eq}(\phi)$ and $b_f(\phi)$ is one of the most remarkable predictions of the present theory, bearing profound physical implications. To see this let us recall that the dimensionless mobility $b_f(\phi)$ is just the  scaled long-time self-diffusion coefficient $ D^*(\phi,T_f)\equiv D_L(\phi,T_f)/D^0$ of the fully equilibrated system at the final point $(\phi,T_f)$, which for the present isochoric quench down to zero temperature, $T_f=0$, is the dimensionless equilibrium long-time self-diffusion coefficient of the hard sphere liquid, $ D^*_{HS}(\phi)\equiv  D^*(\phi,T_f=0)$.  This property can be calculated using the equilibrium version of the present theory \cite{todos1} and, as discussed below (see fig. \ref{fig9}(b)), such calculation leads to the prediction that $D^*_{HS}(\phi)$ vanishes at $\phi^{(a)}=0.582$, according to the power law $ D^*_{HS}(\phi)\propto (\phi^{(a)}-\phi)^{2.2}$. As a consequence, if $t^{eq}(\phi) \approx 0.25 \times b_f^{-1}(\phi)\ (\propto D^{*-1}_{HS}(\phi))$, we must expect that as $\phi \to \phi^{(a)}$ the equilibration time will diverge according to  $t^{eq}(\phi) \propto (\phi^{(a)}-\phi)^{-2.2}$.

This predicted divergence of the equilibration time constitutes a
strong and interesting proposal, which  requires, of course, a
critical assessment and validation. We shall return to this
discussion later on in the paper, but at this point, let us carry
out a similar analysis, now using the $\alpha$-relaxation time $
\tau_{\alpha}(t;\phi)$ (whenever we omit the wave-vector $k$ as
argument of $ \tau_{\alpha}(k,t;\phi)$ is because a specific value
for $k$ is being assumed fixed, most frequently $k\approx
k_{max}$). Thus, in Fig.  \ref{fig8}(b) we plot $
\tau_{\alpha}(t;\phi)$ as a function of $t$ for the same set of
values of the volume fraction $\phi$ as in Fig. \ref{fig8}(a).
Here again we find that $ \tau_{\alpha}(t;\phi)$ reaches its
equilibrium value $ \tau^{eq}_{\alpha}(\phi)$   after the same
evolution time $t^{eq}(\phi)$ as in the case of $b(t;\phi)$. Also
here the solid circles highlight the points $(t^{eq}(\phi),
\tau^{eq}_{\alpha}(\phi))$, and the dashed line of the figure
imply that $t^{eq}(\phi) \approx 40 \times
[\tau^{eq}_{\alpha}]^{0.95}(\phi)$. This implies that the time
$t^{eq}(\phi)$ required to equilibrate the system will grow at
least about as fast as the equilibrium value $\tau^{eq}_{\alpha}$ of the
$\alpha$-relaxation time, and that both properties increase
strongly with $\phi$.

Although one can discuss additional features of the class of irreversible process corresponding to the full  isochoric equilibration of the system after its sudden cooling, it is now important to contrast the scenario just described, with that of the second class of irreversible processes. This involves the dynamic arrest of the system, and is the subject of the following section.

\section{Aging of soft-sphere liquids.}\label{sectionV}

Let us recall at this point that the NE-SCGLE description of the spontaneous evolution of the structure and dynamics of an \emph{instantaneously} and \emph{homogeneously} quenched liquid  is provided by the simultaneous solution of Ecs. (\ref{relsigmadif2pp})-(\ref{fluctsquench}). As discussed in subsection \ref{subsectionIII.2}, there exist two fundamentally different classes of irreversible isochoric processes, represented by the vertical downward arrows in Fig. \ref{fig1}. In the previous section we described the resulting scenario for the most familiar of them,  namely, the full isochoric equilibration of the system. In this section we present the NE-SCGLE description of the second class of irreversible isochoric processes, in which the system starts in an ergodic state and ends in the region where it is expected to become dynamically arrested.

Thus, let us continue considering the TLJ model system introduced in Sect. \ref{sectionIII} (Eq. \ref{truncatedlj}, with $\nu=6$), subjected to the sudden isochoric cooling at fixed volume fraction $\phi=0.6$, larger than  $\phi^{(a)}=0.582$, from the point $(\phi,T^{*(i)}=0.1)$ in the ergodic region, to the point $(\phi,T^{*(f)}=0)$ in the region of dynamically arrested states. By construction, the solution $\gamma^{(i)}$ of Eq. (\ref{nep5pp}), obtained using $S^{(eq)}(k;\phi,T^{*(i)})$ as the structural input, is $\gamma^{(i)}=\infty$. In this sense, the present class of process is identical to the first one, discussed in the previous section. The main difference lies, of course, in the fact that in the present case the solution  of Eq. (\ref{nep5pp}) for the squared localization length $\gamma^{(f)}$, obtained using the equilibrium static structure factor $S^{(eq)}(k;\phi,T^{*(f)})$ of the final point as input, will now have a \emph{finite} value.

To see the consequences of this difference, let us go back to subsection \ref{subsectionIII.3}, and consider the function $S^*(k;u)$ in Eq. (\ref{solsdktexp}), with $0\le u \le \infty$. For each value of $u$ we may use $S^*(k;u)$  in the  bifurcation equation (\ref{nep5pp}) for  $\gamma(t)$, now denoted as $\gamma^*(u)$. Throughout the previous section it was implicitly assumed that $\gamma^*(u)=\infty$ for $0\le u \le \infty$, an assumption based on the fact that the system started and ended in a fluid-like state. In the present case, however, although the system starts with the condition that $\gamma^*(u=0)= \infty$, we know that the final point $(\phi,T^{*(f)}=0)$ corresponds to an arrested state, so that $\gamma^{(f)}\equiv \gamma^*(u=\infty)$ has a finite value. This means that somewhere between $u=0$ and $u=\infty$ the function $\gamma^*(u)$ changed from infinity to a finite value, and this then implies the existence of  a finite value  $u^{(a)}$ of $u$, such that $\gamma^*(u)$ remains infinite only within the interval $0\le u < u^{(a)}$. Thus, in the present case the simultaneous solution of Ecs. (\ref{relsigmadif2pp})-(\ref{fluctsquench}) starts in practice with the precise determination of  $u^{(a)}$.

\subsection{Method of solution of Ecs. (\ref{relsigmadif2pp}), (\ref{bdt})-(\ref{fluctsquench}) for aging.}\label{subsectionV.1}

To determine the critical value $u^{(a)}$, let us consider again the sequence $S^*(k;u_n)$ of snapshots of the static structure factor, generated by the expression in Eq. (\ref{solsdktexp}) with  $u_n=n\Delta u$ ($n=0,1,2,...$). Since we have assumed that initially the system is fluid-like, the value of $u^{(a)}$ cannot be $u^{(a)}=0$. Thus, let us employ each snapshot of the sequence $S^*(k;u_n)$, with $n=1,2,...$, as the static input of   Eq. (\ref{nep5pp}), thus determining the sequence
$\gamma^*_n \equiv \gamma^*(u_n)$ of values of $\gamma^*(u)$, which
starts with $\gamma^*_0=\infty$. If $\gamma^*_1$ turns out to be
finite, then one may take a smaller $u$-step $\Delta u$, until this
does not happen. For a sufficiently small $\Delta u$, there will be
an integer $n_a$  such that $\gamma^*_n=\infty$ for $n < n_a$ and
$\gamma^*_n$ is finite for $n > n_a$, i.e., such that $u_{n_a}<
u^{(a)} < u_{(n_a+1)}$. This process can be refined by decreasing
$\Delta u$, so that one can determine $u^{(a)}$ with arbitrary
precision for the given initial and final conditions
$(\phi,T^{*(i)})$ and $(\phi,T^{*(f)})$. For example, one can readily perform this procedure for the quench indicated by the right arrow of Fig. \ref{fig1} (from the point $(\phi,T^{*(i)}=0.1)$ to the final point $(\phi,T^{*(f)}=0)$ at fixed $\phi=0.6$),  with the result  $u^{(a)}$= 0.0128.

\begin{figure}
\begin{center}
\includegraphics[scale=.23]{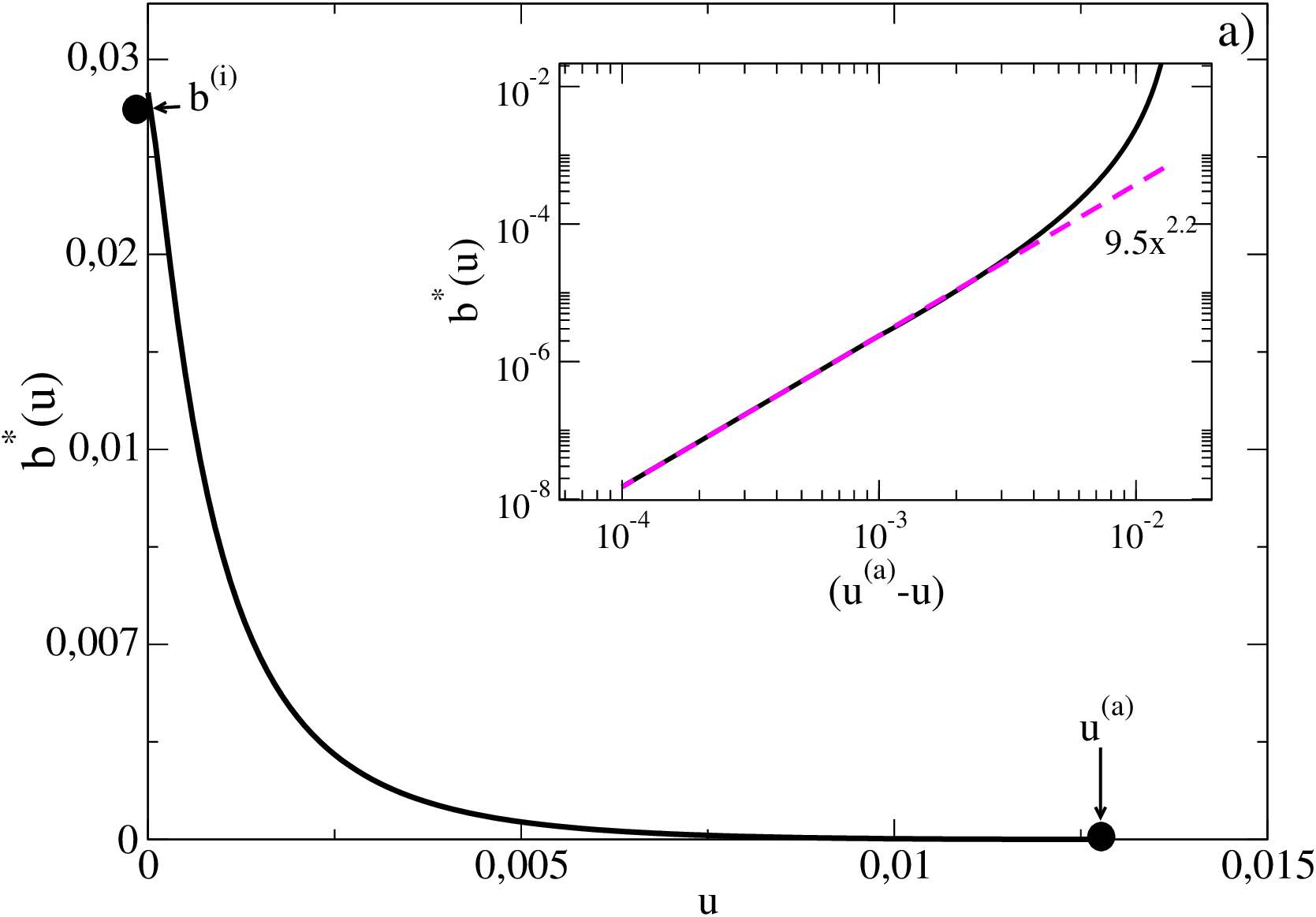}
\includegraphics[scale=.25]{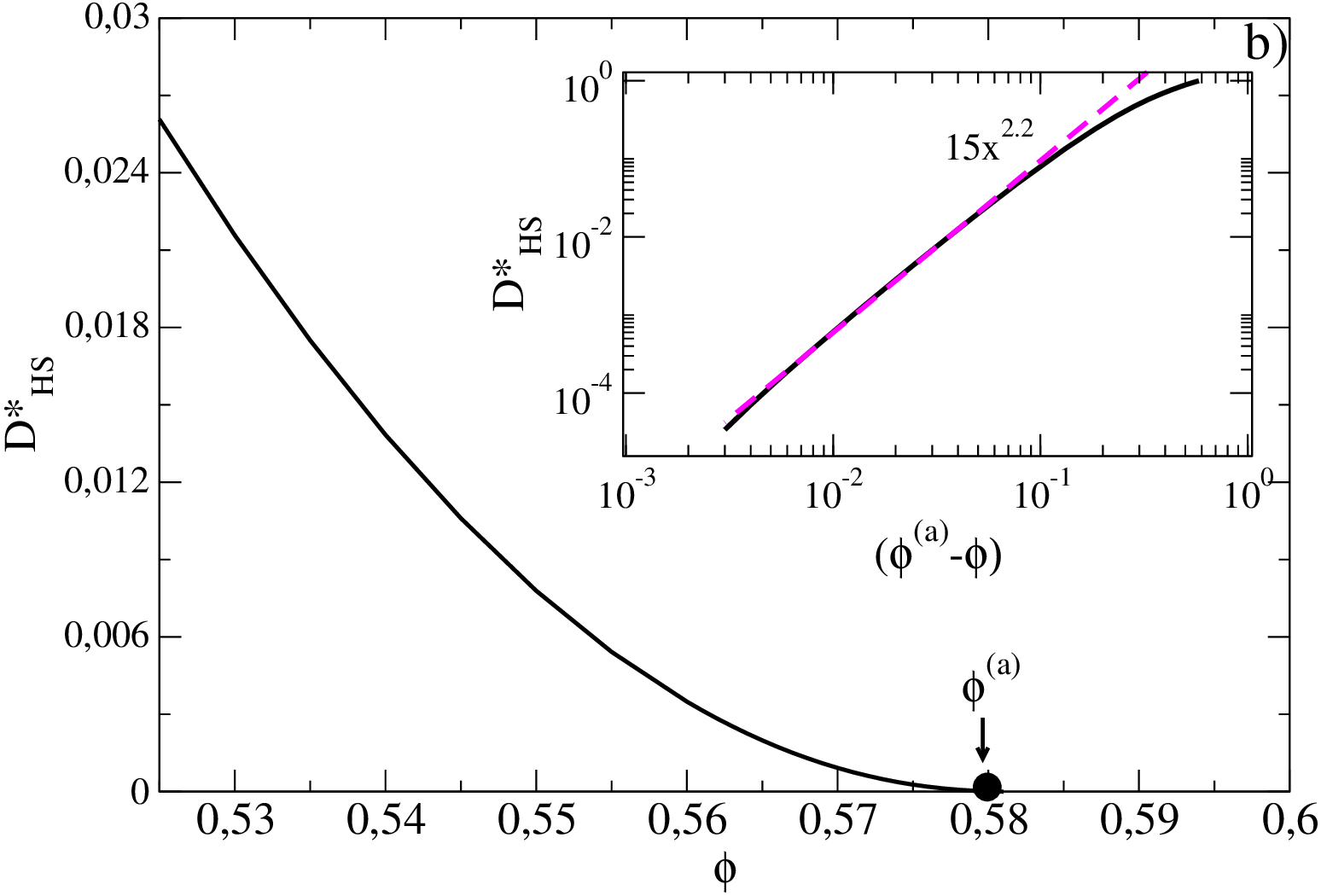}
\caption{(a) Mobility function $b^*(u)$ plotted as a function of the parameter $u$ and of the  difference
$(u^{(a)}-u)$ (inset) for the sudden isochoric cooling from the point $(\phi=0.6,T^{*(i)}=0.1)$  to the point $(\phi=0.6,T^{*(f)}=0)$. (b) Scaled long-time self-diffusion coefficient $ D^*_{HS}(\phi)\equiv D_L(\phi)/D^0$ of the hard-sphere liquid as a function of volume fraction $\phi$ and of the difference $(\phi^{(a)}-\phi)$ (inset).} \label{fig9}
\end{center}
\end{figure}

Once one has determined $u^{(a)}$ with the desired
precision, one can construct a new sequence $u_l$ of $(N+1)$
equally-spaced values of $u$, defined as $u_l\equiv l\times
(u^{(a)}/N)$ with $0 \le l \le N$, along with the corresponding
sequence $S^*(k;u_l)$ of snapshots of the static structure factor
(using Eq. (\ref{solsdktexp})). Since the sequence $S^*(k;u_l)$ is
identical to the sequence $S(k;t_l)$ (with $t_l$ such that
$u_l=\int_0^{t_l} b(t')dt'$), to each member of this sequence,  the
self-consistent Eqs. (\ref{dzdtquench})-(\ref{fluctsquench}) assigns
a snapshot of the full dynamics of the system. In particular, the
use of Eq. (\ref{bdt}) generates a sequence of values $b^*(u_l)$ of
the mobility $b^*(u)$, with arbitrary resolution (set by the number
$N$ of $u$-steps). To illustrate these concepts, in Fig.
\ref{fig9}(a) we present the results for $b^*(u)$ corresponding to
the specific quench under discussion. Notice that, as
expected, $b^*(u)\to 0$ as $u$ approaches $u^{(a)}$ from below.

A simple ansatz to model this limiting behavior is
\begin{equation}
b^*(u) \approx B_0 (u^{(a)}-u)^{\beta}. \label{bstardu}
\end{equation}
In the inset of Fig. \ref{fig9}(a) we plot $b^*(u)$ vs. $(u^{(a)}-u)$ to determine the value of the exponent $\beta$ and  the pre-factor $B_0$, with the result $\beta$= 2.2 and $B_0$= 9.5. We performed similar calculations varying the initial temperature $T^{*(i)}$, and found the value $\beta$= 2.2 of the exponent is independent of $T^{*(i)}$, so that the dependence of $b^*(u)$ on the initial temperature is carried only in the pre-factor  $B_0$. For example, we found that $B_0(T^{*(i)})=\ 9.5,\ 33$, and 490, for $T^{*(i)}$= 0.1, 0.05, and 0.01, respectively.

Another remarkable feature of the $u$-dependence of $b^*(u)$ illustrated in the inset of Fig.
\ref{fig9}(a)  is its similarity with the volume fraction dependence of the scaled long-time self-diffusion coefficient of the  \emph{fully equilibrated}  hard-sphere system, $ D^*_{HS}(\phi)\equiv D_L(\phi,T_f=0)/D^0$. This property can be calculated using the equilibrium version of the SCGLE theory \cite{todos1}, and the results are exhibited in Fig. \ref{fig9}(b). As discussed before \cite{todos2}, the theoretical prediction is that $D^*_{HS}(\phi)$ vanishes at the dynamic arrest volume fraction $\phi^{(a)}=0.582$. The results of Fig. \ref{fig9}(b) show that in the vicinity of $\phi^{(a)}$, the function $D^*_{HS}(\phi)$ follows the power law $ D^*_{HS}(\phi)\propto (\phi^{(a)}-\phi)^{2.2}$, i.e., it vanishes at $\phi^{(a)}$ with the same exponent as $b^*(u)$ vanishes at $u=u^{(a)}$.

\subsubsection{Asymptotic decay  $b(t)\propto t^{-\eta}$.}\label{subsubsectionV.1.1}

\begin{figure}
\includegraphics[scale=.23]{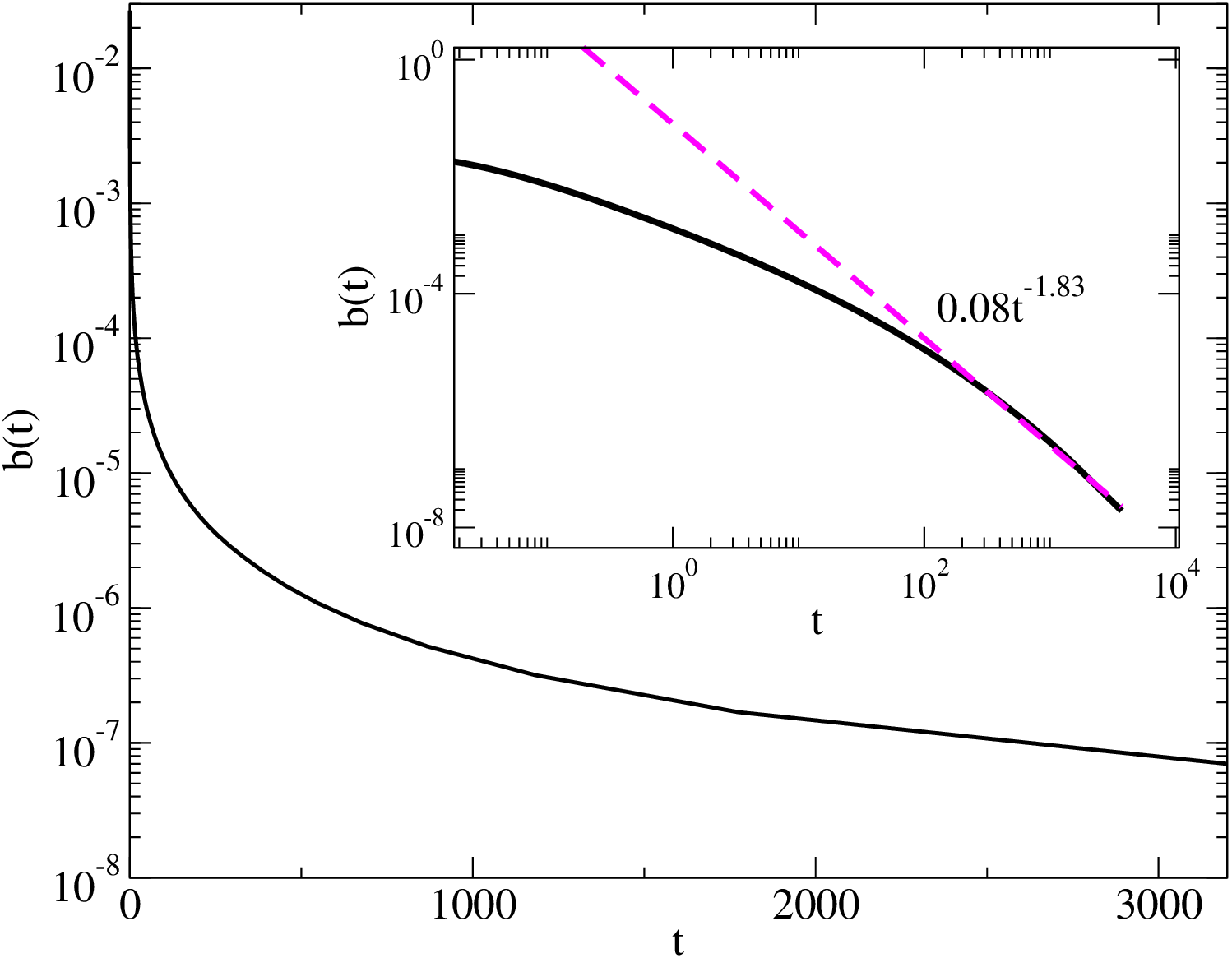}
\caption{Non-equilibrium mobility $b(t)$ as a function of evolution time $t$ for the quench processes from the point $(\phi=0.6,T^{*(i)}=0.1)$  to the point $(\phi=0.6,T^{*(f)}=0)$. The inset exhibits the long-$t$ asymptotic
decay $b(t) \approx b_0 t^{-\eta}$ described by Eq. (\ref{bdtasympt}).} \label{fig10}
\end{figure}

At this point let us notice that the sequence $b^*(u_l)$ must be
identical to the sequence $b(t_l)$ of values of $b(t)$ at the times $t_l\equiv \int_0^{u_l} [1/b^*(u')]du'$. The sequence of times $t_l$ can be determined by
means of the approximate recurrence relationship in Eq. (\ref{recrel0}), i.e.,
\begin{equation}\nonumber
t_{l+1}=t_l + \Delta u/ b^*(u_l), \label{recrelp}
\end{equation}
with $\Delta u\equiv u^{(a)}/N$, and it allows us to transform the
sequence $b^*(u_l)$ into the discrete representation $b(t_l)$ of the
function $b(t)$. The results for $b(t)$  are plotted in Fig.
\ref{fig10}, to exhibit the fundamentally different behavior of
the functions $b^*(u)$ and $b(t)$. While the former has a
well-defined zero at a finite value of its argument, namely, at
$u=u^{(a)}$, the function $b(t)$ decays to zero in a much slower
fashion. In fact, as we now discuss, one of the main predictions of
the NE-SCGLE theory is that $b(t)$ will remain finite for any finite
time $t$, and only at $t=\infty$ the mobility will reach its
asymptotic value of zero. Thus, the system in principle will always
remain fluid-like, and the dynamic arrest condition $b(t)=0$ will
only be reached after an infinite waiting time.

Let us actually demonstrate that the value of $t$ corresponding to
$u^{(a)}$ is $t^{(a)}=\infty$, and that the mobility $b(t)$ decays
as a power law with $t$. To discuss the first issue, let us recall
Eq. (\ref{tdu00}), which writes the function $u(t)$ as
\begin{equation}
t(u)=\int_0^u \frac{du'}{b^*(u')}, \label{tdup}
\end{equation}
where the function $b^*(u)$ is, of course, $b^*(u)=b(t(u))$.
According to this result, and to Eq. (\ref{bstardu}), we can write
\begin{equation}
t(u)-t(u_0)=\int_{u_0}^u du'/B_0 (u^{(a)}-u')^{\beta} =
\frac{(u^{(a)}-u)^{-(\beta-1)}}{(\beta-1)B_0}-
\frac{(u^{(a)}-u_0)^{-(\beta-1)}}{(\beta-1)B_0} \label{tdu0}
\end{equation}
for $u$ in some vicinity $u_0\le u\le u^{(a)}$ of $u^{(a)}$. This
implies that, if the exponent $\beta$ is larger than unity, then
$t(u)$ will diverge  as $u$ approaches  $ u^{(a)}$ according to
\begin{equation}
t(u) \approx \frac{(u^{(a)}-u)^{-(\beta-1)}}{(\beta-1)B_0}.
\label{tduasympt}
\end{equation}
As a consequence, the dynamic arrest time $t^{(a)}\equiv t(u^{(a)})$
will be infinite, which is what we set out to demonstrate.

Let us now discuss the possibility that $b(t)$ decays as a power law
with $t$. For this, let us invert the function $t(u)$ in the
previous equation, and write it as
\begin{equation}
u(t)\approx u^{(a)}- \{ (\beta-1)B_0 t \}^{-\frac{1}{(\beta-1)} }.
\label{udtasympt}
\end{equation}
Since, according to Eq. (\ref{udt}), $b(t) = du(t)/dt$, the time
derivative of this asymptotic expression will yield the asymptotic
form for $b(t)$, namely,
\begin{equation}
b(t)\approx b_0 t ^{-\eta }, \label{bdtasympt}
\end{equation}
with
\begin{equation}
b_0 \equiv  [ (\beta-1)^\beta B_0  ]^{-\frac{1}{(\beta-1)} } ,
\label{bmincero}
\end{equation}
and
\begin{equation}
\eta \equiv \frac{\beta}{(\beta-1)} \ \ \ \ \ \ \ \left( \textrm{or}
\ \  (\eta-1) = \frac{1}{(\beta-1)} \right) . \label{etadebeta}
\end{equation}

The latter result implies that if one of the exponents ($\beta$ or
$\eta$) is larger than unity, then the other is also larger than
unity. It also implies that if one of them is larger than 2, then
the other is smaller than 2, and viceversa. In the inset of Fig. \ref{fig10}
we compare the actual NE-SCGLE results for $b(t)$ in the main figure, with the approximate asymptotic expression in Eq. (\ref{bdtasympt}) with a fitted exponent $\eta$, with the result that $\eta=1.83$. This value coincides with the expected result  $\eta =\beta/(\beta-1)$ with $\beta =2.2$. As indicated above, we performed similar calculations varying the initial temperature $T^{*(i)}$, and found that the scenario just described is indeed independent of $T^{*(i)}$. Thus, in the asymptotic expression in Eq. (\ref{bdtasympt}) only the pre-factor  $b_0$ depends on  $T^{*(i)}$, and the approximate expression in  Eq. (\ref{bmincero}) provides an indicative estimate of its actual value.

\subsubsection{Dynamically arrested evolution of $S(k;t)$.}\label{subsubsectionV.1.2}

The properties of the non-equilibrium mobility function $b(t)$ that we have just described reveals the main feature of the time evolution of the static structure factor $S(k;t)$ when the system is driven to a point $(\phi,T^*)$ in the region of dynamically arrested states. We refer to the fact that under such conditions, the long-time asymptotic limit of  $S(k;t)$ will no longer be the expected equilibrium static structure factor $S^{(eq)}(k;\phi,T^*)$, but another,  well-defined non-equilibrium static structure factor  $S^{(a)}(k)$, given by
\begin{equation}
S^{(a)}(k)=S^{(i)}(k)e^{-\alpha (k)u^{(a)}}+ S^{eq}_f(k) \left(1-e^{-\alpha (k)
u^{(a)}}\right).  \label{nesadk}
\end{equation}
This non-equilibrium static structure factor not only depends on the final point $(\phi,T^*)$, but also on the protocol of the quench (in the present instantaneous isochoric quench, this means on the initial temperature $T^{*(i)}$).

To see the emergence of this scenario, let us consider the sequence $S^*(k;u_l)$ of snapshots of the static structure factor generated with Eq. (\ref{solsdktexp}), for the finite sequence $u_l$ of $(N+1)$ equally-spaced values of $u$ defined as $u_l\equiv l\times (u^{(a)}/N)$ with $0 \le l \le N$. According to Eq. (\ref{tdup}), and to its asymptotic version in Eq. (\ref{tduasympt}), in the present case the finite range  $0\le u\le u^{(a)}$ maps onto the infinite physically relevant range $0\le t \le \infty$ of the evolution time $t$ (in contrast with the equilibration processes studied in the previous section, in which the infinite range  $0\le u\le \infty$  maps onto the infinite range $0\le t \le \infty$). Since the sequence $S^*(k;u_l)$ is identical to the sequence $S(k;t_l)$, with $t_l=\int_0^{u_l} du'/b^*(u')$,  then the sequence of snapshots  $S(k;t_l)$ describing the full evolution of  $S(k;t)$ will be generated by a sequence of snapshots of  $S^*(k;u)$ with $u$ only in the range $0\le u\le u^{(a)}$. In other words, in the present case none of the snapshots  of  $S^*(k;u)$ with  $u\ge u^{(a)}$ will map onto any  physically observable snapshot of $S(k;t)$, and this applies in particular to the snapshot $S^*(k;u=\infty)$, corresponding to  the expected equilibrium static structure factor $S^{(eq)}(k)$. In this manner, the long time limit of $S(k;t)$, normally being the ordinary equilibrium value $S^{(eq)}(k)\ (\equiv [\overline{n}\mathcal{E}^{(f)}(k)]^{-1})$, is now replaced by a non-equilibrium dynamically arrested static structure factor $S^{(a)}(k)$ given, according to Eq. (\ref{solsdktexp}), by the expression in Eq. (\ref{nesadk}).

Besides the remarkable prediction of the existence of this well-defined non-equilibrium asymptotic limit of $S(k;t)$, the second relevant feature refers to the kinetics of $S(k;t)$ as it approaches $S^{(a)}(k)$. To exhibit this feature, let us subtract Eq. (\ref{nesadk}) from Eq. (\ref{solsigmadkt}). This
leads to
\begin{equation}
S(k;t)-S^{(a)}(k) = A(k) \left[e^{-\alpha
(k)[u(t)-u^{(a)}]}-1\right], \label{sdklong}
\end{equation}
with
\begin{equation}
A(k) \equiv e^{-\alpha (k)u^{(a)}}\{S^{(i)}(k)-S^{eq}_f(k)
\}. \label{defadk}
\end{equation}
At long times, when $[u(t)-u^{(a)}]$ is small, this equation reads
\begin{equation} S(k;t)-S^{(a)}(k)  \approx
A(k)\alpha(k)[u^{(a)}-u(t)]. \label{sdklonglong}
\end{equation}
From Eq. (\ref{udtasympt}), however, we have that
$u^{(a)}-u(t)\approx  \{ (\beta-1)B_0 t \}^{-\frac{1}{(\beta-1)} }$,
so that the previous long-time expression for $S(k;t)$ can be
written as
\begin{equation} S(k;t)-S^{(a)}(k)  \approx D(k)
 t^{-\frac{1}{(\beta-1)} }, \label{sdklonglong}
\end{equation}
with
\begin{equation} D(k) \equiv
A(k)\alpha(k)\{ (\beta-1)B_0  \}^{-\frac{1}{(\beta-1)} }.
\label{sdklonglong}
\end{equation}

\begin{figure}
\begin{center}
\includegraphics[scale=.3]{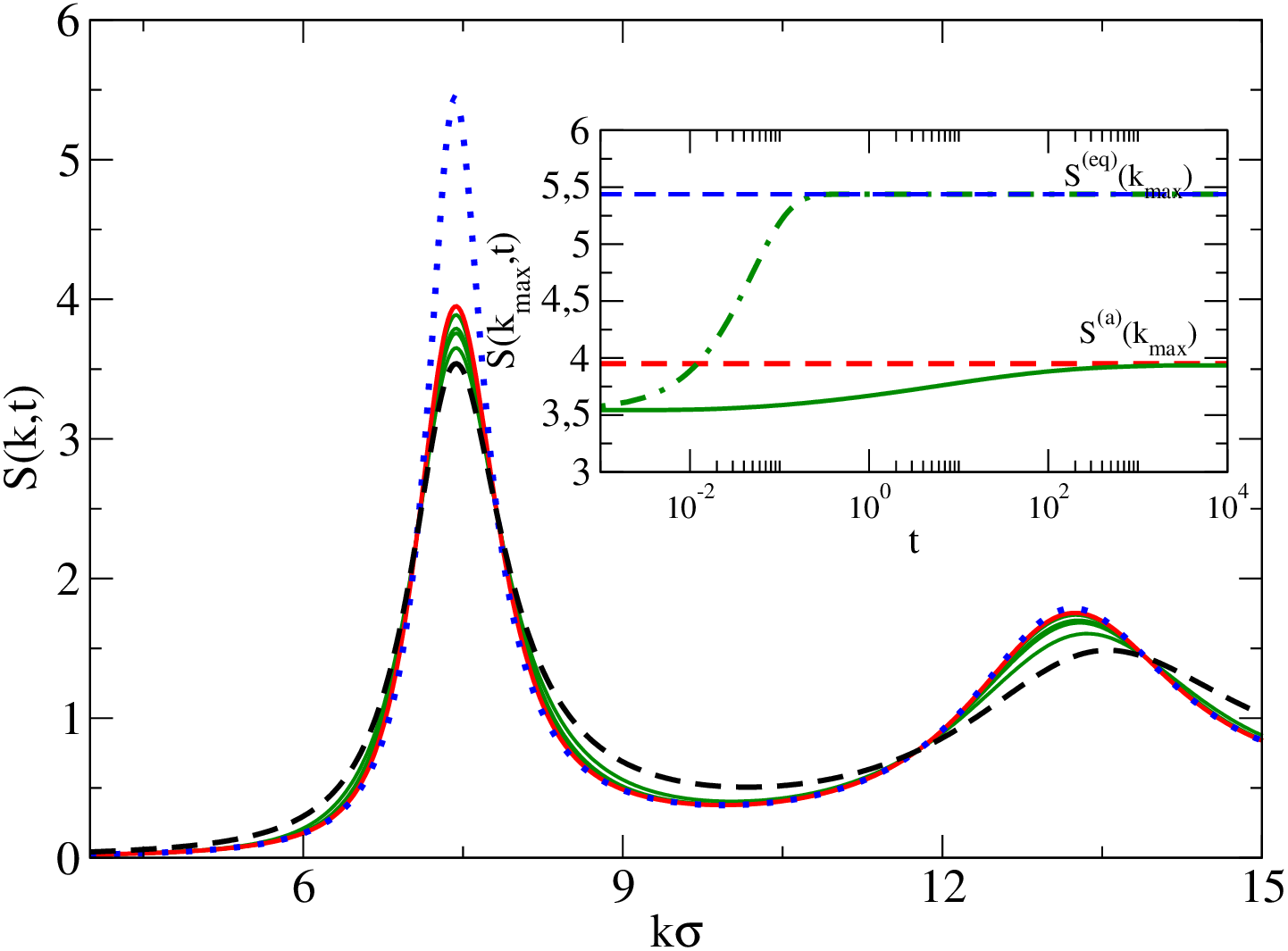}
\caption{Snapshots of the non-equilibrium evolution of  $S(k;t)$ (thin solid (green) lines) corresponding to the isochoric quench at fixed volume fraction $\phi=0.6$, from $T^{*(i)}=0.1$ to $T^{*(f)} =0$. The dashed (black) line is the initial structure factor $S^{(i)}(k)$. The dotted (blue) line is $S^{eq}_f(k)$. The thick solid (red) line is the dynamically arrested asymptotic solution $S^{a}(k)$ of Eq. (\ref{relsigmadif2pp}), given by Eq. (\ref{nesadk}). In the inset, the solid line is the maximum of $S(k;t)$ as a function of the evolution time $t$, and the dashed line is the maximum of $[S^{(i)}(k)e^{-\alpha (k)b^{(i)}t}+ [\overline{n}\mathcal{E}^{(f)}(k)]^{-1}(1-e^{-\alpha (k)b^{(i)}
t})]$.} \label{fig11}
\end{center}
\end{figure}

Thus, we conclude that, contrary to the kinetics of the equilibration process, in which  $S(k;t)$ approaches $S^{(eq)}(k)$ in an exponential-like fashion, this time the decay of $S(k;t)$ to its stationary value $S^{(a)}(k)$ follows a power law. At very  short times, however, $b(t) \approx b^{(i)}$, and hence, $u(t)\approx b^{(i)}t$. Thus, according to Eqs. (\ref{solsigmadkt}) and (\ref{udt}), we have that the very initial evolution of $S(k;t)$ might seem to approach its expected equilibrium value $S^{(eq)}(k) =[\overline{n}\mathcal{E}^{(f)}(k)]^{-1}$ in an apparently ``exponential"  manner, with a relaxation time $t^{app} \approx 1/ \alpha(k) b^{(i)}$. This apparent initial exponential evolution, however, crosses over very soon to the much slower long-time evolution of $S(k;t)$ described by the  asymptotic expression in Eq. (\ref{sdklonglong}).

Fig. \ref{fig11} illustrates with  a sequence of snapshots the predicted non-equilibrium evolution of $S(k;t)$ after the isochoric quench at $\phi=0.6$ from $T^{*(i)}=0.1$ to $T^{*(f)}=0$. There we highlight the initial static structure factor $S^{(i)}(k)=S^{(eq)}(k;\phi,T^{*(i)})$ and the dynamically arrested long-time asymptotic limit $S^{(a)}(k)$ of the  non-equilibrium evolution of $S(k;t)$. For reference, we also plot the expected, but inaccessible, equilibrium static structure factor $S^{(eq)}(k;\phi,T^{*(f)}) =1/\overline{n}\mathcal{E}(k;\phi,T^{*(f)})$ corresponding to the final temperature $T^{*(f)}=0$. Regarding the kinetics of the non-equilibrium evolution, in the inset we  plot the evolution of the maximum of $S(k;t)$ as a function of $t$ to illustrate the fact that $S(k;t)$ approaches $S^{(a)}(k)$ much more slowly, in fact as the power law $[S^{(a)}(k)-S(k;t)]\propto t^{-0.83}$. For reference, we also plot the maximum of the function $[S^{(i)}(k)e^{-\alpha (k)b^{(i)}t}+ [\overline{n}\mathcal{E}^{(f)}(k)]^{-1}(1-e^{-\alpha (k)b^{(i)}
t})]$ which, according to  Eq. (\ref{solsdktexp}), would describe the evolution of  $S(k;t)$ if $b(t)$ remained constant,  $b(t)=b^{(i)}$.

\subsection{Aging of the dynamics}\label{subsectionV.2}

\begin{figure}
\begin{center}
\includegraphics[scale=.25]{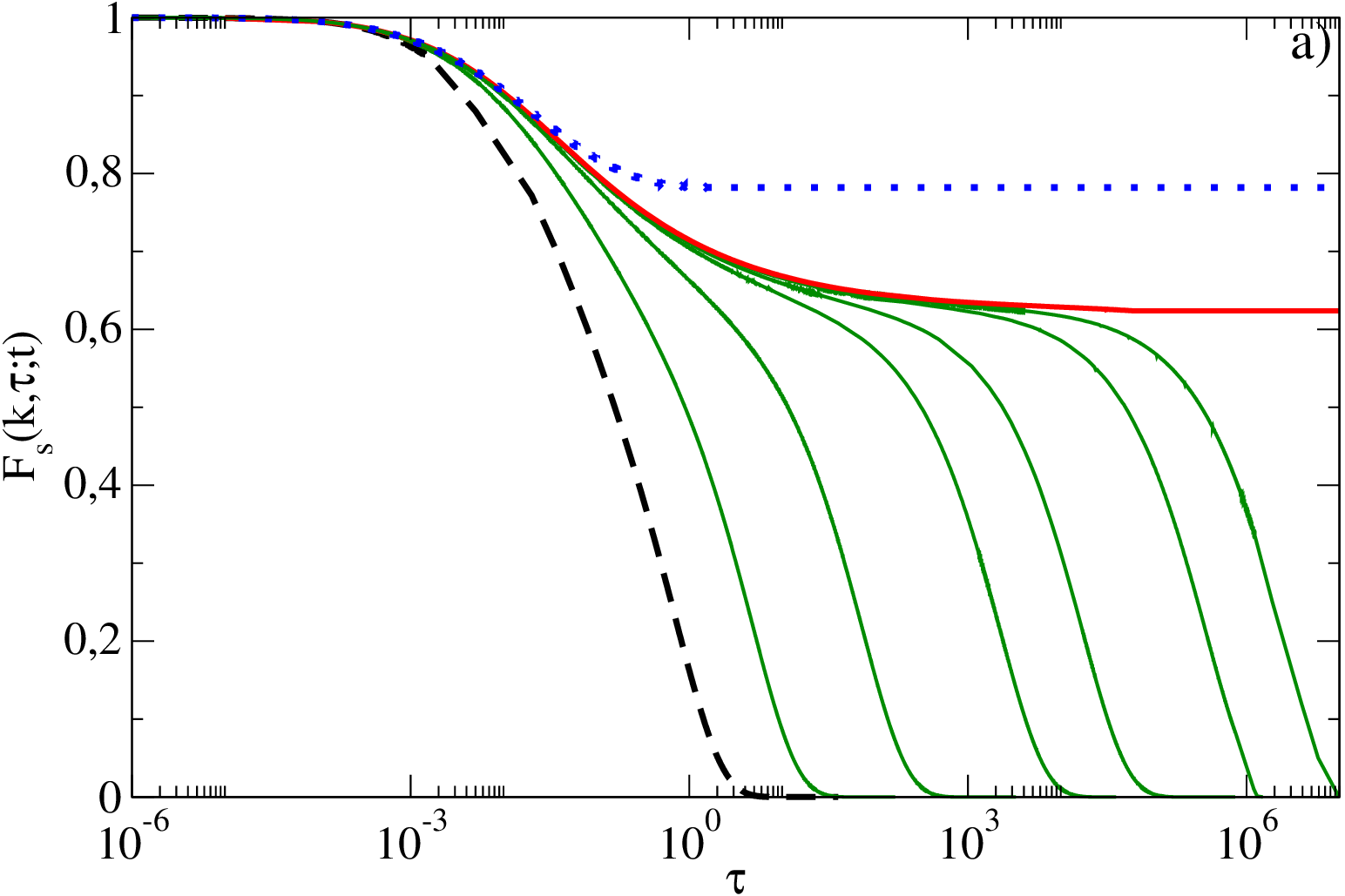}\hskip1cm
\includegraphics[scale=.27]{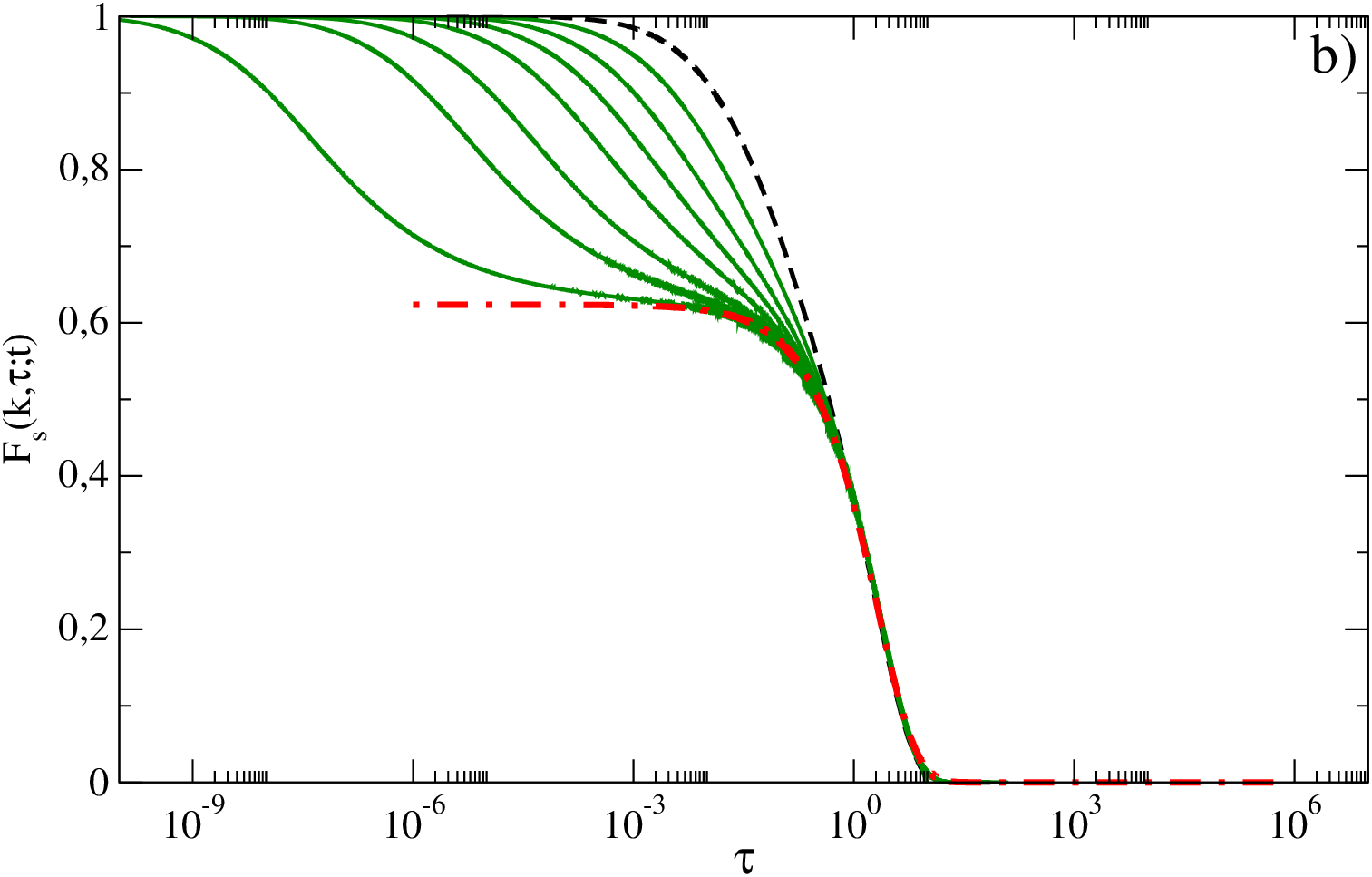}
\caption{(a) Sequence of snapshots of the intermediate scattering function $F_S(k,\tau;t)$ at $k=7.1$ (thin solid (green) lines) plotted as a function of correlation time $\tau$ for a sequence of values of the  waiting time $t$ ($=\ 0.25,\ 5.6,\ 106,\ 400,\ 590,$ and 1600) after the sudden temperature quench at fixed volume fraction $\phi=0.6$, from $T^{*(i)}=0.1$ to $T^{*(f)} =0$. The dashed line is the initial ISF $F_S(k,\tau;t=0)$, the thick solid (red) line is the non-equilibrium asymptotic limit $F_S^{(a)}(k,\tau)\equiv F_S(k,\tau;t=\infty)$, and the dotted (blue) line is the expected (but inaccessible) equilibrium ISF corresponding to the hard-sphere system at $\phi=0.6$. (b) Same sequence of snapshots of $F_S(k,\tau;t)$ plotted as a function of the time $t$ scaled with the $\alpha$-relaxation time $\tau_\alpha(t)$. The dot-dashed line here is the stretched exponential $0.624\times \exp {[-0.528(t/\tau_\alpha)^{0.82}]}$ } \label{fig12}
\end{center}
\end{figure}

Let us now discuss how the scenario just described manifests itself in the non-equilibrium evolution of the dynamics. We first recall that for each snapshot of the static structure factor $S(k;t)$, the solution of  Eqs. (\ref{bdt})-(\ref{lambdadk}) determines a snapshot at waiting time $t$ of each of the dynamic properties of the system. Thus, the process of dynamic arrest may also be observed, for example, in terms of the $t$-evolution of the self  intermediate scattering function $F_S(k,\tau;t)$ or of the $\alpha$-relaxation time $\tau_{\alpha}(k;t)$. In Fig. \ref{fig12}(a) we present a sequence of snapshots of the ISF $F_S(k,\tau;t)$ (thin solid lines), evaluated at the fixed wave-vector $k=7.1$, plotted as a function of correlation time $\tau$, for a sequence of waiting times $t$ after the sudden temperature quench from $T^{*(i)}=0.1$ to $T^{*(f)} =0$ at fixed volume fraction $\phi=0.6$.

In the figure we highlight with the dashed line the initial ISF $F_S(k,\tau;t=0)$. The (arrested) non-equilibrium asymptotic limit $F_S^{(a)}(k,\tau)\equiv \lim_{t\to \infty}F_S(k,\tau;t)$ is indicated by the solid line, whereas the dotted line denotes the inaccessible equilibrium ISF $F_S^{(eq)}(k,\tau)$,  i.e., the solution of Eqs. (\ref{dzdtquench})-(\ref{lambdadk}) in which the final \emph{equilibrium} static structure factor  $S^{(eq)}(k;\phi=0.6,T^{*(f)}=0)$ (also inaccessible) is employed as static input. We observe that at $t=0$, $F_S(k,\tau;t)$ shows no trace of dynamic arrest, but as the waiting time $t$ increases, its relaxation time increases as well. In the figure we had to stop at a finite waiting time, but the theory predicts that the ISF $F_S(k,\tau;t)$ will always decay to zero for any finite waiting time $t$, and continues to evolve forever, yielding always a finite,  ever-increasing, $\alpha$-relaxation time $\tau_\alpha (k;t)$. The relaxation of $F_S(k,\tau;t)$ is characterized by a fast initial decay ($\beta$-relaxation) to an increasingly better defined plateau, whose height $f_0(k)$ is not determined by the expected equilibrium ISF $F^{eq}_S(k,\tau)$, but by the non-equilibrium asymptotic limit $F_S^{(a)}(k,\tau)$. In other words, $f_0(k)$ is the ``true" non-equilibrium non-ergodicity parameter $f_S^{(a)}(k) \equiv \lim_{\tau\to\infty}F_S^{(a)}(k,\tau)$.

From this sequence of snapshots of $F_S(k,\tau_{\alpha};t)$ we can extract the $t$-evolution of the  $\alpha$-relaxation time $\tau_\alpha (k;t)$ defined in Eq. (\ref{taualphadef}). The results allows us to notice one of the main features of the predicted long-$\tau$ decay of $F_S(k,\tau;t)$, namely,  the long-time collapse of the curves representing  $F_S(k,\tau;t)$, corresponding to different evolution times $t$ (like those in Fig.  \ref{fig12}(a)), onto the same stretched-exponential curve upon scaling the correlation time $\tau$ with the corresponding  $\tau_{\alpha}(k;t)$. In other words, at long times  $F_S(k,\tau;t)$ scales as
\begin{equation}
F_S(k,\tau;t)\approx f_0 e^{-a_0\left(\frac{\tau}{\tau_{\alpha}(k;t)}\right)^{\beta}}, \label{stretched1}
\end{equation}
where $f_0$ is the height of the plateau of $F_S^{(a)}(k,\tau)$ and $a_0=1+ \ln f_0$ (so that $F_S(k,\tau_{\alpha};t)=e^{-1}$), and with $\beta$ being a fitting parameter. This scaling is illustrated in Fig.  \ref{fig12}(b) with the sequence of results for  $F_S(k,\tau;t)$ in Fig.  \ref{fig12}(a) now plotted in this scaled manner, which are then well represented by the stretched-exponential function above, with $f_0=0.624$, $a_0=0.528$ and $\beta=0.82$.

\begin{figure}
\begin{center}
\includegraphics[scale=.25]{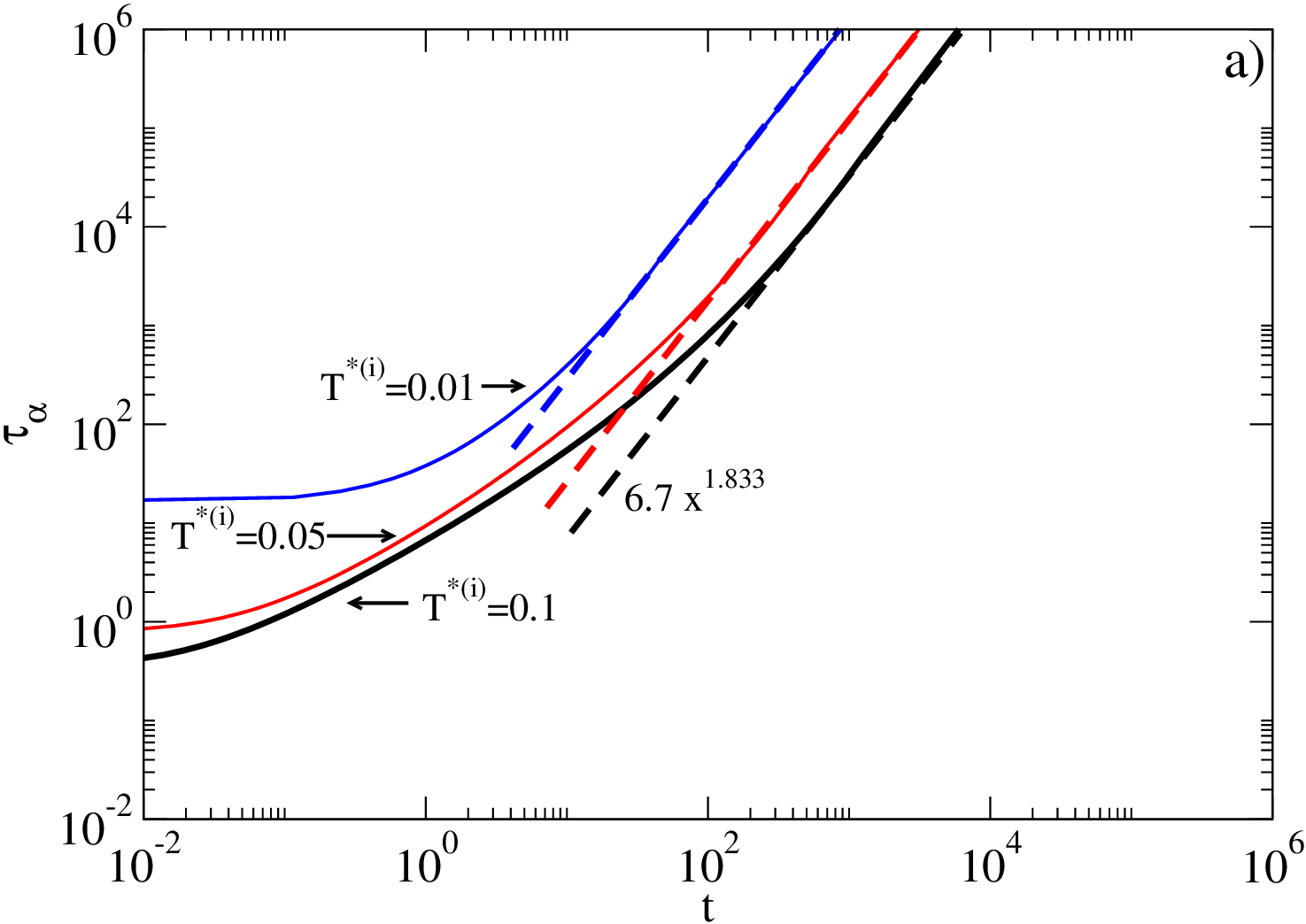}
\includegraphics[scale=.25]{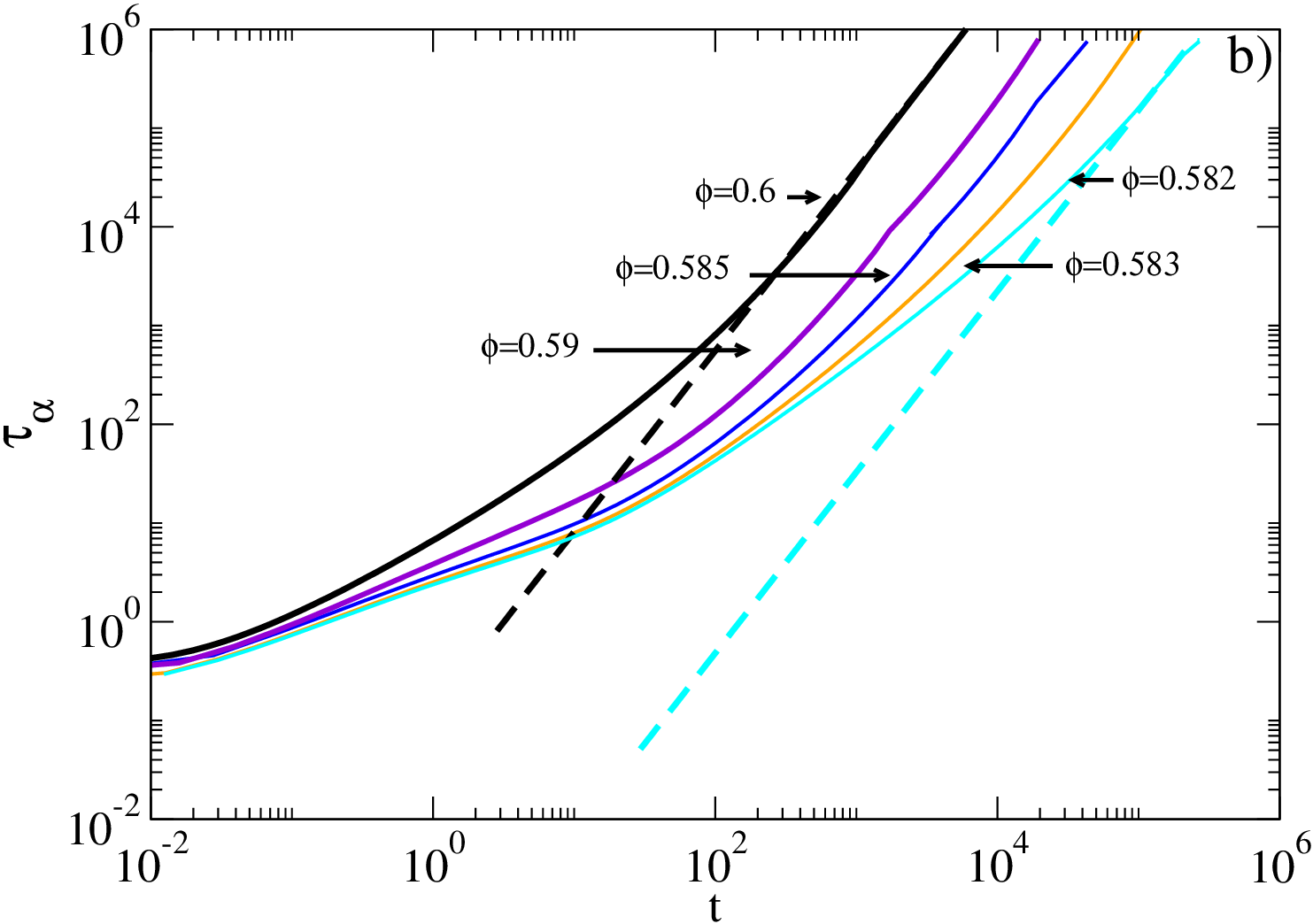}
\caption{Waiting-time dependence of the $\alpha$-relaxation time $\tau_\alpha(t;T^{*(i)},\phi)$ (defined in the text) after the sudden temperature quench at fixed volume fraction $\phi$, from an initial temperature $T^{*(i)}$ to a final temperature $T^{*(f)} =0$. In (a) we present the results for the initial temperatures $T^{*(i)}=0.1$, 0.05, and 0.001 at the same volume fraction $\phi=0.6$. In (b) we fix the initial temperature $T^{*(i)}=0.1$ and present results for $\phi=0.6$ and other volume fractions. The dashed lines indicate the asymptotic power-law $\tau_\alpha(t)\propto A t^x$ that fits the results in the indicated regimes.} \label{fig13}
\end{center}
\end{figure}

The non-equilibrium evolution of the dynamics can be summarized by plotting $\tau_\alpha (k;t)$ as a function of waiting time $t$. This is done here in Fig. \ref{fig13}, where we plot $\tau_\alpha (t)\ (\equiv \tau_\alpha (k=7.1,t))$ as a function of $t$. The thick dark solid line in Fig. \ref{fig13}(a) and (b) derive from the sequence of snapshots of $F_S(k,\tau_{\alpha};t)$ in Fig. \ref{fig12}(a), corresponding to the quench at $\phi=0.6$ with initial temperature $T^{*(i)}=0.1$. As indicated in these figures, at long waiting times we find that $\tau_{\alpha}(k;t)$ increases with $t$ according to a power law that is numerically indistinguishable from $\tau_{\alpha}(t)\propto t^{\eta}$ with $\eta\approx 1.83$. In other words the present theory predicts, taking into account Eq. (\ref{bdtasympt}), that at long waiting times, $\tau_{\alpha}(k;t)$ diverges with $t$ with the same power law as $b^{-1}(t)$.

Besides these results, Fig. \ref{fig13}(a) also presents theoretical results for two additional quench programs that differ only in the initial temperature, namely, $T^{*(i)}=0.05$ and $T^{*(i)}=0.01$. These three initial temperatures lie above the dynamic arrest transition temperature $T^{*(a)}(\phi)$ corresponding to the isochore $\phi=0.6$, which is $T^{*(a)}(\phi=0.6)= 0.004$. The first feature to notice is that the detailed waiting time dependence of $\tau_\alpha$ at short times may be strongly
quench-dependent, but the asymptotic power law $\tau_{\alpha}(t)\propto t^{\eta}$ with the exponent $\eta\approx 1.83$ is independent of the initial temperature  $T^{*(i)}$. Complementing this information, Fig.  \ref{fig13}(b) describes the dependence of the evolution of $\tau_{\alpha}(t;T^{*(i)},\phi)$ on the value of the volume fraction $\phi$ at which these isochoric processes occur, assuming that each of them start and end at the same initial and final temperatures, $T^{*(i)}=0.1$ and $T^{*(f)}=0$. The main feature to notice in these results is that the long-time asymptotic growth of  $\tau_{\alpha}(t;\phi)$ with waiting time $t$ is also characterized by the power law  $\tau_{\alpha}(t)\propto t^{1.83}$.

\section{Crossover from equilibration to aging.}\label{sectionVI}

\begin{figure}
\includegraphics[scale=.25]{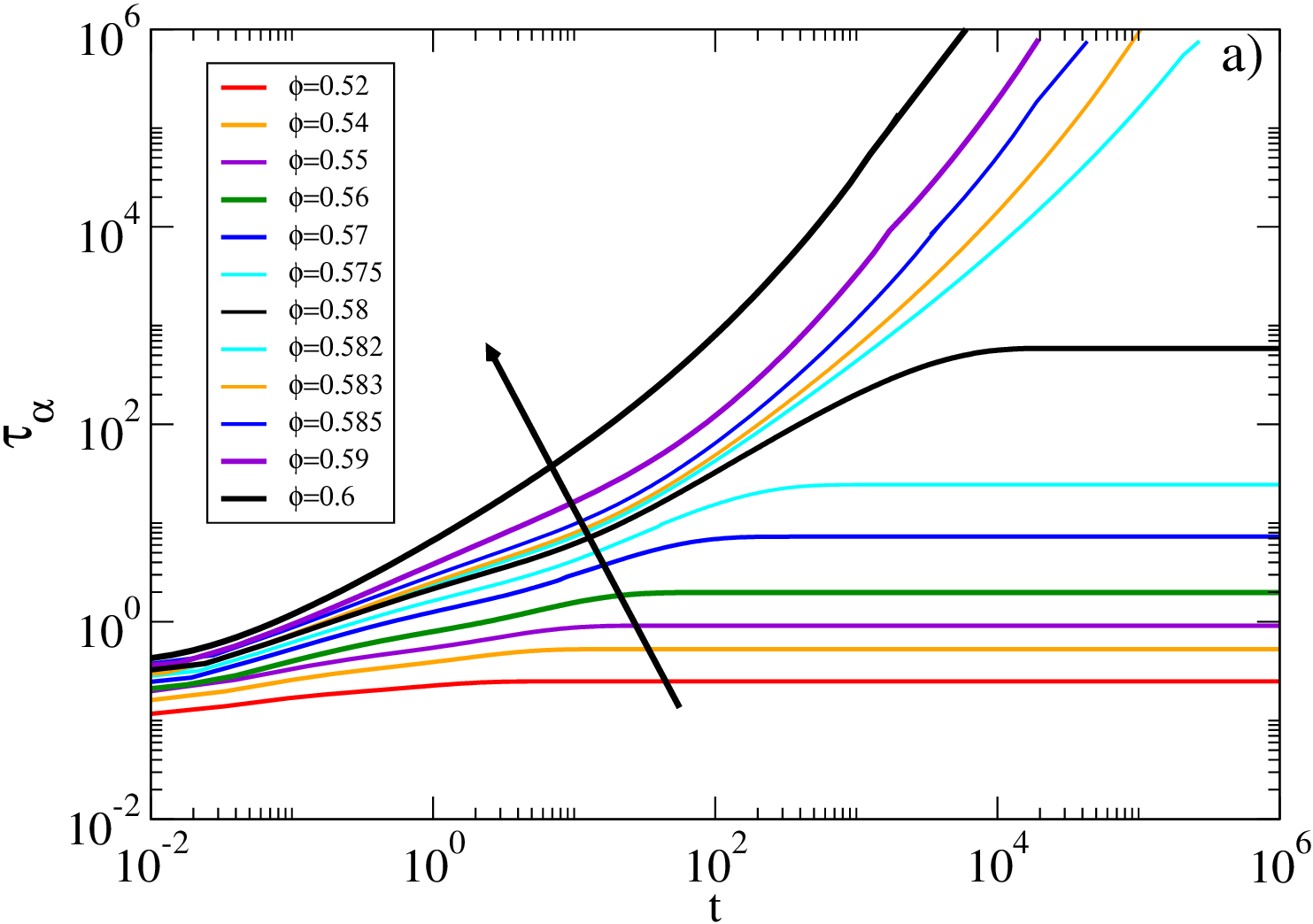}
\includegraphics[scale=.25]{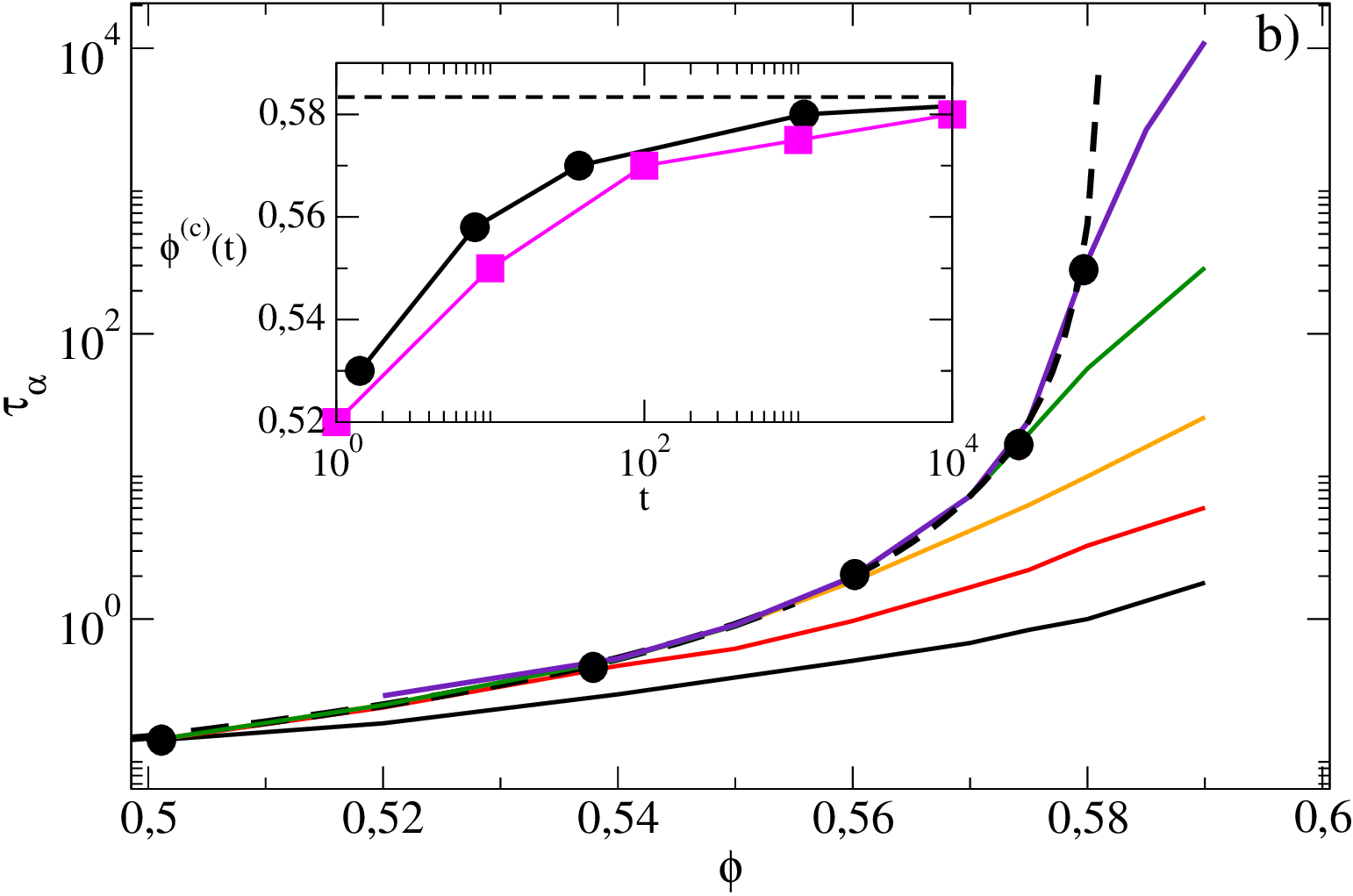}
\caption{(a) Waiting time dependence of the $\alpha$-relaxation
time $\tau_{\alpha}(t;\phi)$ for a sequence of fixed volume fractions (the arrow indicates increasing $\phi$).
(b) $\phi$-dependence of the $\alpha$-relaxation time
$\tau_{\alpha}(t;\phi)$ for the sequence of fixed waiting times $t = 10^0,\
10^1,\ 10^2,\ 10^3,\ \textrm{and }\ 10^4$ (from bottom to top).  The dashed line is $\tau_{\alpha}^{eq}(\phi)\equiv \lim_{t \to \infty}\tau_{\alpha}(t;\phi)$, which is the \emph{equilibrium} $\alpha$-relaxation
time of the hard-sphere system, predicted by the equilibrium SCGLE theory. The solid circles highlight the crossover points
$(\phi^{(c)}(t),\tau_{\alpha}^{eq}(\phi^{(c)}))$ for each waiting time $t$ shown. The inset of (b) shows the
evolution of the crossover volume fraction $\phi^{(c)}(t)$ predicted by the NE-SCGLE theory (solid line) and determined in
the simulations of Ref. \cite{gabriel} (solid squares). } \label{fig14}
\end{figure}

Of course, one could continue describing the predictions of the NE-SCGLE theory regarding the detailed evolution of each relevant structural and dynamic property of the glass-forming system along  the process of equilibration or aging. At this point, however,  we would like to unite the main results of the previous two sections in a single integrated scenario that provides a more vivid physical picture of the predictions of the present theory. With this intention, in Fig.  \ref{fig14}(a) we have put together the results for the isochoric evolution of $\tau_{\alpha}(t;\phi)$ previously  presented in Figs. \ref{fig8}(b) and \ref{fig13}(b), corresponding to the quench at fixed volume fraction $\phi$, from an initial temperature $T^{*(i)}=0.1$ to a final temperature $T^{*(f)} =0$,  for volume fractions $\phi$ smaller and larger than $\phi^{(a)}=0.582$.

Displaying together these results allows us to have a richer and more comprehensive scenario of the transition from equilibration processes to aging processes in the soft-sphere glass-forming liquid, discussed separately in the previous two sections. According to the NE-SCGLE theory, the dynamic arrest transition is in principle a discontinuous transition, involving the abrupt passage from one pattern of evolution (equilibration) to the other (aging) when the control parameter $\phi$ crosses the singular value $\phi^{(a)}=0.582$. The discontinuous nature of this kinetic transition is rooted in the abrupt transition contained in the \emph{equilibrium} version of the SCGLE  theory, which actually predicts the existence and location of the dynamic arrest transition line (see Fig.  \ref{fig13}). The zero-temperature limit of this transition line corresponds to the critical volume fraction $\phi^{(a)}=0.582$. Thus, the evolution of the system after the temperature quench from an initial temperature $T^{*(i)}=0.1$ to a final temperature $T^{*(f)}=0$ is dramatically different if the volume fraction of the isochoric process is smaller or larger than this critical volume fraction.

However, in order to actually witness this dramatic difference, we would have to perform observations at volume fractions infinitesimally closer to $\phi^{(a)}$, and within an evolution time window much larger (in fact, infinite) than in  any real experiment or simulation. In fact, what we would like to illustrate now is that the experimental observation of the consequences of this theoretically-predicted singularity will be blurred by this unavoidable finiteness of the time window of any experimental observation. To see this, let us display the same information presented in Fig.  \ref{fig14}(a), which plots $\tau_{\alpha}(t;\phi)$ as a function of $t$ for a sequence of volume fractions $\phi$, in a complementary format. This is done in Fig.  \ref{fig14}(b), which plots $\tau_{\alpha}(t;\phi)$ as a function of $\phi$ for a sequence of waiting times $t$.

The main feature to notice in each of the curves corresponding to a fixed waiting time $t$, is that one can distinguish two regimes in volume fraction, namely, the low-$\phi$ (equilibrated) regime and the high-$\phi$ (non-equilibrated) regime, separated in  a continuous fashion, and not as an abrupt transition, by a crossover volume fraction $\phi^{(c)}(t)$. Focusing, for example, on the results corresponding to $t=10^3$, we notice that $\phi^{(c)}(t=10^3)\approx 0.57$. In Fig. \ref{fig14}(b) we have highlighted the crossover points $\phi=\phi^{(c)}(t)$, $\tau_{\alpha}(t,\phi)=\tau^{eq}_{\alpha}(\phi^{(c)})$, corresponding to each waiting time $t$ considered. We observe that the resulting crossover volume fraction $\phi^{(c)}(t)$ first increases rather fast with $t$, but then slows down considerably, reaching a theoretical maximum crossover volume fraction, $\lim_{t\to\infty} \phi^{(c)}(t)$,  given by $\phi^{(a)}=0.582$, as indicated in the inset of the figure.

The scenario illustrated by Fig.  \ref{fig14}(b) has additional physical implications. Although it is impossible to witness the infinite-time implications of the theoretically-predicted singular dynamic arrest transition, it is important to stress that its finite-time consequences, such as those illustrated in this figure, can be predicted, and could be corroborated by performing measurements at \emph{intentionally} finite, accessible waiting times.  It is thus important to test if these predictions make sense by comparing them with available experimental or simulation data. Although this comparison falls out of the scope of the present paper, we can say that the picture that emerges from the predicted dependence of $\tau_{\alpha}(t;\phi)$ on waiting time and volume fraction just discussed, is fully consistent with the most relevant qualitative features observed in a simulation experiment consisting precisely of the equilibration of a hard-sphere liquid, initially prepared in a non-equilibrium state \cite{gabriel}. In fact, such simulation experiment was originally inspired by the very initial version of the theoretical scenario of the present non-equilibrium theory. To have an idea of the level of agreement, in the inset of  Fig.  \ref{fig14}(b) we have included the simulation data for the evolution of the crossover volume fraction $\phi^{(c)}(t)$ with  waiting time reported in Ref \cite{gabriel}. In a separate communication \cite{nescgle4} we shall analyze in detail other aspects of the comparison between the predictions of the NE-SCGLE theory and available simulation results, which indicates a general agreement and exhibits some well-defined limitations of the present non-equilibrium theory.

\section{Concluding remarks}

In summary, in this work we have started the systematic exploration of the predicted NE-SCGLE scenario of the irreversible isochoric evolution of a soft-sphere glass-forming liquid whose temperature is suddenly quenched from its initial value $T^{(i)}$ to a final value $T^{(f)}=0$. As we explained here, the response falls in two mutually exclusive possibilities: either the system will reach its new equilibrium state within an equilibration time $t^{eq}(\phi)$ that depends on the fixed volume fraction $\phi$, or the system ages forever in the process of becoming a glass.

In the first case the equilibrium $\alpha$-relaxation time $\tau_\alpha^{eq}(\phi)$, and the equilibration time $t^{eq}(\phi)$ needed to reach thermodynamic equilibrium, are predicted to remain finite for volume fractions smaller than a critical value $\phi^{(a)}\approx 0.582$, but as  $\phi$ approaches this hard-sphere dynamic-arrest volume fraction, both characteristic times will diverge and will remain infinite for $\phi\ge \phi^{(a)}$. Although it is intrinsically impossible to witness the actual predicted divergence, the theory makes distinct predictions regarding the transient non-equilibrium evolution occurring within experimentally-reasonable waiting times $t$, which could, thus, be compared with realizable experiments or simulations.

This applies even more to the predictions regarding the complementary regime, $\phi \ge \phi^{(a)}$, in which the system, rather than ever reaching equilibrium, is predicted to age forever. As discussed in the previous section, under these circumstances the long-time asymptotic limit of  $S(k;t)$ will no longer be the expected equilibrium static structure factor $S^{(eq)}(k)$, but the non-equilibrium, but well-defined, dynamically arrested static structure factor $S^{(a)}(k)$. Furthermore, $S(k;t)$ is predicted to approach $S^{(a)}(k)$ in a much slower fashion (a power law), in contrast with the exponential-like manner in which $S(k;t)$ approaches $S^{(eq)}(k)$ when the system equilibrates.

Putting together the two regimes just described, we have presented the scenario predicted to emerge for the crossover from equilibration to aging. As discussed in the previous section, the discontinuous and singular behavior is intrinsically unobservable in practice, due to the finiteness of the  time windows of experimental measurements. This forces the discontinuous dynamic arrest transition to appear as a blurred crossover, which may depend on the protocol of the experiment and of the measurements. Testing these predictions by comparing them with available experimental or simulation data is an issue that we shall leave for future studies, since the main purpose here was to provide the details of the methodologies needed to solve the equations that define the NE-SCGLE, and to illustrate its use with the application to the specific system and processes considered here. As indicated at the end of the previous section, we can say that the picture that emerges from the predicted dependence of $\tau_{\alpha}(t;\phi)$ on waiting time and volume fraction is consistent with the most relevant qualitative features observed in the simulation experiment of the equilibration of the hard-sphere liquid \cite{gabriel}. In a separate paper we shall establish a more direct contact with those simulation results, and with other simulation or experimental data.

In the meanwhile, it will also be interesting to interrogate the NE-SCGLE theory on the variations of the scenario just described, when the system and conditions employed here are modified. For example, one may be interested in understanding how this scenario might change  when protocol of the quench is modified. Other questions may refer to the dependence of this scenario on the particular class of model system and interactions (involving here only soft repulsions), particularly when attractive forces are incorporated. The answer to these questions will surely use the methods and experience developed in the presented work, and will be the subject of future research.

\vskip2cm

ACKNOWLEDGMENTS: This work was supported by the Consejo Nacional de
Ciencia y Tecnolog\'{\i}a (CONACYT, M\'{e}xico), through grants
No. 84076 and 132540.

\end{document}